\documentclass[12pt,onecolumn,draftcls,journal]{IEEEtran}
\usepackage{multirow}
\usepackage{sublabel}
\usepackage{amsfonts}
\usepackage{amssymb}

\usepackage{cite}

\ifCLASSINFOpdf
  \usepackage[pdftex]{graphicx}
\else
  \usepackage[dvips]{graphicx}
\fi
\usepackage[cmex10]{amsmath}
\usepackage{array}

\usepackage[tight,footnotesize]{subfigure}

\usepackage[center]{caption}
\begin{document}

%
\title{Self-Iterating Soft Equalizer}





%

\author{\IEEEauthorblockN{Seongwook Jeong, \IEEEmembership{Student Member, IEEE}
and Jaekyun Moon\IEEEauthorrefmark{2}, \IEEEmembership{Fellow,
IEEE}\\}
\IEEEauthorblockA{Dept. of Electrical and Computer Engineering\\
University of Minnesota\\
Minneapolis, Minnesota 55455, U.S.A.\\
Email: jeong030@umn.edu\\}
\IEEEauthorblockA{\IEEEauthorrefmark{2} Dept. of Electrical Engineering\\
Korea Advanced Institute of Science and Technology\\
Daejeon, 305-701, Republic of Korea \\
Email: jmoon@kaist.edu}

\thanks{This work was supported in part by the National Research Foundation of Korea under grant no. 2011-0029854, the KAIST-ICC S\&T Leading Primary Research Program grant no. N10110050 and the NSF under Theoretical Foundation grant no. 0728676 and IHCS grant no. 0701946. This work was presented in part at IEEE Global Telecommunications Conference 2011.}
}


\maketitle
\setlength\arraycolsep{1pt}
\thispagestyle{empty}

\begin{abstract}
A self-iterating soft equalizer (SISE) consisting of a few
relatively weak constituent equalizers is shown to provide robust performance
even in severe intersymbol interference (ISI) channels that exhibit deep nulls and valleys within the signal band. Constituent
equalizers are allowed to exchange soft information in the absence of interleavers
based on the method that are designed to suppress significant correlation among
their soft outputs. The resulting SISE works well as a stand-alone
equalizer or as the equalizer component of a turbo equalization
system. The performance advantages over existing methods are validated with bit-error-rate (BER) simulations and extrinsic information transfer (EXIT) chart analysis. It is shown that in turbo equalizer setting the SISE achieves performance closer to the maximum \textit{a posteriori} probability equalizer than any other known schemes in very severe ISI channels.

\end{abstract}

%

\newpage
\setcounter{page}{1}

\section{Introduction}
Turbo equalization is a well-established technique that is highly
effective in combating intersymbol interference (ISI) via
iterative exchange of soft decisions between a soft-in soft-out
(SISO) equalizer and a SISO error-correction decoder separated by
an interleaver \cite{Turbo}. The Bahl-Cocke-Jelinek-Raviv (BCJR)
algorithm \cite{BCJR74} and the soft-output Viterbi algorithm
(SOVA) \cite{SOVA89} provide excellence performance as the
equalizer component of turbo equalization systems, but both
schemes require implementation complexity that grows exponentially
with the length of the ISI channel.

Numerous suboptimal, low-complexity turbo equalization schemes
have been proposed to mitigate the high computational complexity
of the BCJR and SOVA methods. See, for example, \cite{Wang99, Chan01, Wu01, TE02, TSK02, Laot01, Honig04, Laot05, Sun05, Moon05, Rad05, SFE06, JeongICC10, Jeong10}. 
Some of these schemes utilize reduced-trellis approaches while others rely on filter-based methods such as the linear equalizer (LE) or the decision feedback equalizer (DFE).  
The SISO version of the LE has been discussed in \cite{Wang99, TE02, TSK02, Laot01, Honig04, Laot05, Sun05}. The design is based on applying the classical minimum-mean-squared-error (MMSE) criterion while utilizing second order statistics of the input symbols estimated from the extrinsic information made available by the outer decoder. The authors of \cite{TE02} have shown that using Proakis' well-known example channel with ISI tap weights [1 2 3 2 1], the turbo equalizer based on the SISO LE performs as well as one based on the BCJR equalizer at error rates below $10^{-4}$. While the original formulation of the equalizer in \cite{TSK02}, \cite{TE02} gives rise to a time-varying filter, it has been replaced by a low-complexity quasi-time-invariant filter whose tap setting changes only once in every iteration stage \cite{TSK02}. The low complexity method does not result in a significant performance loss \cite{TSK02}. The same authors also investigated the SISO DFE but concluded that it is inferior to the LE counterpart of \cite{TE02}. The authors of \cite{Moon05} considered a more severe ISI channel of the form [1 2 3 4 3 2 1], a natural extension of the previously considered ISI patterns by Proakis,
and showed that for this channel, there is a substantial performance gap between the LE-based turbo equalizer of \cite{TE02} and the BCJR-based turbo equalizer. They showed that using reduced-trellis search based on soft-decision-feedback can close this performance gap with the BCJR-based turbo equalizer. This approach was also shown to be effective for the magnetic recording channel \cite{Rad05}. 

Another meaningful development on suboptimal equalization is to employ two
DFEs (or DFE variants) running in opposite directions and combine their
extrinsic information \cite{Jeong10, iterBAD03, BiSFE06, BiDFE00, BAD05}. This
``bi-directional" DFE (BiDFE) algorithm takes advantage of
different decision error and noise distributions at the outputs of
the forward and time-reversed DFEs. Moreover, unlike the LE
or the DFE, the BiDFE algorithm can be designed to avoid performance
degradation even when the filter taps are constrained to be
time-invariant \cite{Jeong10}. The BiDFE can be
considered as a parallel concatenated scheme with two suboptimal
DFEs producing somewhat correlated yet significantly different extrinsic information. 
The time-reversal operation applied to the reverse DFE can be viewed as a type of interleaving
that attempts to make two input streams going into the forward and reverse DFEs appear independent.
In \cite{Jeong10}, an effective log-likelihood-ratio (LLR) combining strategy for the two bi-directional DFEs has been proposed. The work of \cite{Jeong10} also showed a new extrinsic formulation that can effectively suppress error propagation in the DFE leading to improved performance of the DFE-based turbo equalizer relative to the LE-based one.
When each DFE employs the extrinsic information formulation method of \cite{Jeong10} and a LLR combining strategy of \cite{Jeong10} is used, the BiDFE is shown to perform considerably better than the LE. 
  
In this paper, we focus on a new equalizer structure that employs the LE, the DFE or the BiDFE as constituent modules, 
devising a strategy that allows iterative exchange of soft information among the constituent equalizers. 
Unlike typical turbo processing methods, no interleaver exists between the SISO equalizer modules,
and a special strategy to combat the correlation between successive module outputs must be devised.
Interleaving in the usual sense is not possible in our case. Placing a shuffling device between two 
component equalizers at the receiver side would imply that two sets of channel output sequences one corresponding to the original input sequence and the other corresponding to the shuffled input sequence are available. 
This would require transmission of a redundant set of data and is clearly not sensible in practice. 
The significant correlation between the equalizer module outputs is a direct consequence of the lack of interleaving. 
It is shown that the extrinsic information of one module becomes the \textit{a priori} information for the next module in concatenation via a specific scaling law that depends on the correlation between the input and output information sets of the first module. This equalizer is viewed as a self-iterating soft equalizer (SISE)
consisting of several suboptimal constituent equalizers which are
concatenated with no interleavers placed between them. 
The rationale behind this particular equalizer structure is that the suboptimal
equalizers such as the LE, the DFE and the BiDFE all have their own
advantages and disadvantages, and one should be able to benefit
from the presence of the other equalizers. For example, the LE
does not have the error propagation problem which the DFE suffers
from, whereas the DFE often shows significantly better performance
than the LE when feedback decisions are correct; and the BiDFE
provides solid performance even with time-invariant filters,
although its complexity is roughly double the complexity of the
DFE. We show that through simulation and analysis this type of 
``self-iteration" among constituent equalizer components with different characteristics 
can improve upon traditional equalization schemes based on a single equalizer component.

The remainder of the paper is organized as follows. In Section
\ref{sec:System_Model}, a brief statement of the system model is
given. In Section \ref{sec:SISE}, we show a proper way to generate \textit{a priori} information from
the extrinsic information out of other constituent equalizers when
the information between the equalizers could be significantly correlated,
and then propose self-iterating soft equalizer design for uncoded
systems. We also provide turbo equalization algorithms based on the
SISE in Section \ref{sec:iterSISE}. In Section \ref{sec:Filters},
we briefly review the individual suboptimal equalizers with the several filter types utilizing \textit{a priori} information which will be employed in the proposed SISE algorithms.  
In Section \ref{sec:Simulation Results},
numerical results and analysis are given. Finally, we draw
conclusions in Section \ref{sec:Conclusion}.

\section{System Model}\label{sec:System_Model}
Given the transmitted sequence of coded bits $\{x_k\}$, the ISI
channel output at time $n$ is
\begin{eqnarray} r_n & = & \sum_{k=0}^{L_h}
{h_k x_{n-k}} + w_n  \label{eq:r_n}
\end{eqnarray}
where $w_n$ is additive white Gaussian noise (AWGN) with variance
$N_0$ and $\{h_k\}$ is the energy-normalized channel impulse
response with length $L_h+1$. In this paper, it is assumed that the
transmitted symbol is a binary input with the average power equal
to 1, i.e., $x_n \in \{ \pm 1 \}$ and $\mathrm{E}(x_n^2) = 1$, and the ISI channel coefficients and noise
samples are real-valued. This restriction is not necessary for the
algorithm development here but makes the presentation simpler. We also assume that the channel response 
is time-invariant and deterministic. The schemes investigated in this paper can be applied to 
random channels, where a channel response changes from one transmission to the next in some random fashion,
provided that the channel remains static over a given transmission period and that channel estimation can be done 
at the receiver side.

In turbo equalization, the \textit{a priori} log-likelihood  ratio
(LLR) of $x_n$ to the equalizer is defined as
\begin{eqnarray}
L_a(x_n) \triangleq \ln \dfrac{\mathrm{Pr}(x_n = +1)}{\mathrm{Pr}(x_n = -1)} \nonumber
\end{eqnarray}
where the probabilities in the expression are in reality just
estimates. The probabilities are all set to 1/2 initially, and then, as the
turbo iteration ensues, the extrinsic information generated by the
outer decoder is used as the \textit{a priori} information to the
equalizer.

Based on these \textit{a priori} LLR values for the symbols, the
equalizer generates its own extrinsic information, which will in
turn be passed to the decoder. Let $y_n$ be the equalizer filter
output sequence corresponding to the observation sequence $r_n$
applied at the input. In an effort to produce the extrinsic
information $L_e(x_n)$ that should not depend on the \textit{a
priori} LLR of the current symbol $x_n$, \textit{$L_a(x_n)$ is set to zero
during the computation of $y_n$} \cite{TE02}. Then, the equalizer's extrinsic
information is directly related to the equalizer output $y_n$ as:
\begin{eqnarray}
L_e(x_n) 
& \triangleq & \ln \dfrac{\mathrm{Pr}(x_n = + 1 \mid y_n)}{\mathrm{Pr}(x_n = - 1 \mid y_n)} \Bigg\vert_{ L_a(x_n) = 0} \nonumber \\
& = & \ln \dfrac{\mathrm{p}(y_n  \mid x_n = + 1)\mathrm{Pr}(x_n = + 1)}{\mathrm{p}(y_n \mid x_n = - 1)\mathrm{Pr}(x_n = - 1)} \Bigg\vert_{ L_a(x_n) = 0} \nonumber \\
& = & \ln \dfrac{\mathrm{p}(y_n  \mid x_n = + 1)}{\mathrm{p}(y_n \mid x_n = - 1)}. \label{eq:Le_xn}
\end{eqnarray}
where $\mathrm{Pr}(A)$ indicates the probability of an event $A$ and $\mathrm{p}(X)$ is the probability density function of a continuous random variable $X$.

\section{Self-Iterating Soft Equalizer Algorithm}\label{sec:SISE}
\begin{figure}[!t]
\centering
\includegraphics[width=10.0cm]{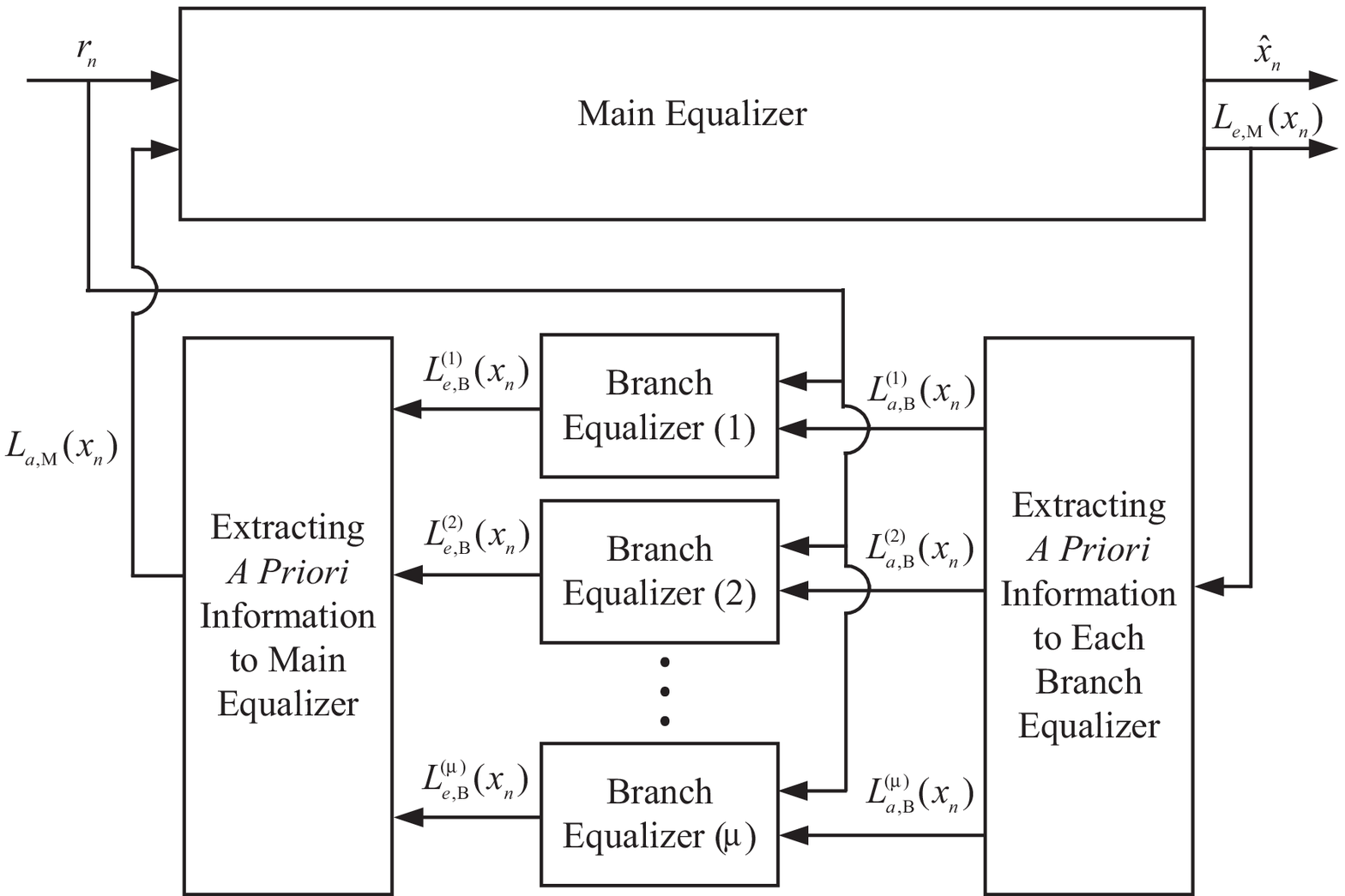}
\caption{Self-iterating soft equalizer.}\label{fig:SISE}
\end{figure}

In this section, we discuss the SISE algorithm. Basically, the SISE we focus on in this paper 
is a SISO equalizer which
consists of one main suboptimal SISO equalizer and $\mu$ branch
suboptimal SISO equalizers. The proposed SISE is illustrated in
Fig. \ref{fig:SISE}.

The key procedure in this algorithm is that the received data
sequence is equalized by the main equalizer and its extrinsic
information is passed to the branch equalizers as their \textit{a
priori} information. The extrinsic information generated in the
branch equalizers is also passed back to the main equalizer to be
used as its \textit{a priori} information for the next stage. Note
that since this equalization algorithm can perform iteratively
without the decoder (hence the name ``self-iterating" equalizer),
it can be used in uncoded systems as well.
The terms ``main" and ``branch" here do not necessarily imply the difference in the complexity levels 
or computational powers between the constituent equalizers. Rather, the distinction simply 
indicates the scheduling strategy, i.e., the main equalizer is the one that makes the initial decision
in the serial concatenation of the constituent equalizers. 
In fact, different arrangements of the 
constituent equalizers are possible, including full parallel concatenation, full serial concatenation and 
combined parallel/serial concatenation, along with many different scheduling strategies.
For the particular SISE structure shown in Fig. \ref{fig:SISE}, for example, it can be seen that while there is ``self-iteration" between the main equalizer and the block of branch equalizers, no self-iterations are assumed among the branch equalizers. While the concept and methods developed in this paper are general,
for the performance analysis and simulation results to be presented, we shall focus on a serial concatenation of 
one main equalizer and one branch equalizer.

Unlike the extrinsic information between the decoder and
the equalizer in usual turbo equalization, the extrinsic
information between the main equalizer and the branch equalizers
have correlation because no interleaving techniques can be used
and their equalization processes are all based on the common
received data sequence. Again, note that placing a shuffler between 
two component equalizers at the receiver side would imply that there is a redundant set of received samples available
corresponding to the shuffled input sequence. This represents a highly wasteful system and would not be practical.
It has been suggested that the high
correlation between the \textit{a priori} information and the
extrinsic information of a module even in a turbo system can cause the
performance degradation \cite{Papke96}, \cite{Huang06}. In this
section, we show a proper way to construct the \textit{a priori} information for the main equalizer 
based on the extrinsic
information generated by the branch equalizers when their outputs are
correlated with the main equalizer output. The same method can be
applied in processing the extrinsic information out of the main
equalizer to obtain the \textit{a priori} information for the branch equalizers
when the main equalizer's soft output is correlated with those of the
branch equalizers.

\subsection{Generation of Uncorrelated \textit{A Priori} Information}
First let us assume that there is one main equalizer and one
branch equalizer in an uncoded system. We will later extend the
proposed algorithm to the case of multiple branch equalizers. We
assume that the main equalizer generates the extrinsic LLR sequence,
$L_{e,\textrm{M}}(x_n)$, which is used to generate the
\textit{a priori} LLR sequence, $L_{a,\textrm{B}}(x_n)$, to be used by the branch equalizer. The branch
equalizer in turn produces its own extrinsic LLR sequence, $L_{e, \textrm{B}}(x_n)$, with
the given $L_{a,\textrm{B}}(x_n)$ sequence. In typical iterative processing,
the $L_{e,\textrm{M}}(x_n)$ sequence simply becomes the $L_{a,\textrm{B}}(x_n)$ sequence (after interleaving) and the $L_{e,\textrm{B}}(x_n)$ sequence obtained based on the $L_{a,\textrm{B}}(x_n)$ sequence and channel observation, in turn, forms the $L_{a,\textrm{M}}(x_n)$ sequence after proper deinterleaving. In the problem at hand, no interleaving/deinterleaving is allowed and there may be significant correlation between the two sequences,
$L_{a,\textrm{B}}(x_n)$ and $L_{e,\textrm{B}}(x_n)$. The question is how we should generate $L_{a,\textrm{M}}(x_n)$ to pass onto the main equalizer, given this correlation. The answer turns out to be a specific scaling law between $L_{a,\textrm{B}}(x_n)$ and $L_{e,\textrm{B}}(x_n)$, as already described in \cite{JeongGC11}. Below, we provide an improved derivation/justification 
for the same scaling law. 

Modeling the \textit{a priori} LLR and the extrinsic LLR
as the output of an equivalent AWGN channel
\cite{Brink01}, we start by writing the unbiased versions of these LLRs associated with the branch equalizer 
as the transmitted symbol $x$
corrupted by AWGN:
\begin{eqnarray}
Y_{a,\textrm{B}}  =  x + u_{a,\textrm{B}} , \qquad
Y_{e,\textrm{B}}  =  x + u_{e,\textrm{B}} \nonumber
\end{eqnarray}
where time index $n$ is dropped for notational simplicity; the process
remains identical as $n$ evolves. The noise terms $u_{a,\textrm{B}}$
and $u_{e,\textrm{B}}$ are assumed to be zero-mean Gaussian random
variables which are independent of the transmitted data $x$ but
correlated with each other with correlation coefficient $\rho_B$.
Note that $Y_{e,\textrm{B}}$ for a specific time point is obtained with the given sequence of $Y_{a,\textrm{B}}$ samples. 

If $Y_{a,\textrm{B}}$ and $Y_{e,\textrm{B}}$ are uncorrelated, the \textit{a priori} information to the main equalizer should be given as $L_{a, \textrm{M}}(x) = L_{e, \textrm{B}}(x) = \ln \left\{ {\textrm{p} \left( Y_{e, \textrm{B}} |  x=+1 \right)}/{\textrm{p} \left(  Y_{e, \textrm{B}} |  x=-1 \right)} \right\}$. However, since the two outputs are correlated and only the extrinsic information should be fed back to the main equalizer, the \textit{a priori} information to the main equalizer can be defined as
\begin{eqnarray}
L_{a, \textrm{M}}(x) & \triangleq & \ln {\dfrac{\textrm{p} \left( Y_{e, \textrm{B}} , Y_{a, \textrm{B}} | x=+1 \right)}{\textrm{p} \left(  Y_{e, \textrm{B}} , Y_{a, \textrm{B}} | x=-1 \right)}} - \ln {\dfrac{\textrm{p} \left(  Y_{a, \textrm{B}} | x=+1 \right)}{\textrm{p} \left(   Y_{a, \textrm{B}} | x=-1 \right)}}  \label{eq:LaM} 
\end{eqnarray}
which reduces to $L_{a, \textrm{M}}(x) = \ln \left\{ {\textrm{p} \left( Y_{e, \textrm{B}} |  x=+1 \right)}/{\textrm{p} \left(  Y_{e, \textrm{B}} |  x=-1 \right)} \right\}$ when $Y_{e, \textrm{B}}$ and $Y_{a, \textrm{B}}$ are independent. The first term in (\ref{eq:LaM}) signifies that 
both $Y_{a,\textrm{B}}$ and $Y_{e,\textrm{B}}$ may contain useful information that can be passed onto the main equalizer, whereas the subtraction of the second term is needed to suppress the information that has originated from the main equalizer itself.
Replacing the likelihood function involving the correlated signals $Y_{a,\textrm{B}}$ and $Y_{e,\textrm{B}}$ with the likelihood function associated with their whitened versions, 
we can further write:
\begin{eqnarray}
L_{a, \textrm{M}}(x) & = & \ln {\dfrac{\textrm{p} \left( Y'_{e, \textrm{B}} , Y'_{a, \textrm{B}} | x=+1  \right)}{\textrm{p} \left( Y'_{e, \textrm{B}} , Y'_{a, \textrm{B}} | x=-1  \right)}} - \ln {\dfrac{\textrm{p} \left(  Y_{a, \textrm{B}} | x=+1 \right)}{\textrm{p} \left(   Y_{a, \textrm{B}} | x=-1 \right)}} \nonumber   \\
& = & \ln {\dfrac{\textrm{p} \left( Y'_{e, \textrm{B}} | x=+1  \right)}{\textrm{p} \left( Y'_{e, \textrm{B}} | x=-1  \right)}} + \ln {\dfrac{\textrm{p} \left( Y'_{a, \textrm{B}} | x=+1  \right)}{\textrm{p} \left( Y'_{a, \textrm{B}} | x=-1  \right)}} - \ln {\dfrac{\textrm{p} \left(  Y_{a, \textrm{B}} | x=+1 \right)}{\textrm{p} \left(   Y_{a, \textrm{B}} | x=-1 \right)}}\nonumber \\
& = & \dfrac{ \left( N_{a,\textrm{B}} - \rho_B \sqrt{N_{a,\textrm{B}} N_{e,\textrm{B}}} \right) } { \left(1-\rho_B^2 \right) N_{a,\textrm{B}} } L_{e,\textrm{B}}(x) + \dfrac{ \left( N_{e,\textrm{B}} - \rho_B \sqrt{N_{e,\textrm{B}} N_{a,\textrm{B}}} \right) } { \left(1-\rho_B^2 \right) N_{e,\textrm{B}} } L_{a,\textrm{B}}(x) - L_{a,\textrm{B}}(x)  \label{eq:apost_branch2}  \\
& = & \dfrac{ \left( N_{a,\textrm{B}} - \rho_B \sqrt{N_{a,\textrm{B}} N_{e,\textrm{B}}} \right) } { \left(1-\rho_B^2 \right) N_{a,\textrm{B}} } L_{e,\textrm{B}}(x) + \dfrac{ \left(  \rho_B^2 N_{e,\textrm{B}} - \rho_B \sqrt{N_{e,\textrm{B}} N_{a,\textrm{B}}} \right) } { \left(1-\rho_B^2 \right) N_{e,\textrm{B}} } L_{a,\textrm{B}}(x)  \label{eq:apost_branch3} 
\end{eqnarray}
where $N_{a,\textrm{B}} \triangleq \mathrm{Var}(u_{a,\textrm{B}})$ and
$N_{e,\textrm{B}} \triangleq \mathrm{Var}(u_{e,\textrm{B}})$, and $Y'_{a,\textrm{B}}$ and $Y'_{e,\textrm{B}}$ are the whitened versions of $Y_{a,\textrm{B}}$ and $Y_{e,\textrm{B}}$. Let $\mathbf{R}$ 
be the noise
correlation matrix:
\begin{eqnarray}
\mathbf{R} & \triangleq & \left[ {\begin{array}{*{20}c}
   {\mathrm{Var}(u_{a,\textrm{B}}) } & {\mathrm{E}(u_{a,\textrm{B}} u_{e,\textrm{B}}) }  \\
   {\mathrm{E}(u_{e,\textrm{B}} u_{a,\textrm{B}}) } & {\mathrm{Var}(u_{e,\textrm{B}}) }  \\
\end{array}} \right]  =  \left[ {\begin{array}{*{20}c}
   N_{a,\textrm{B}}  & {\rho_B \sqrt{ N_{a,\textrm{B}} N_{e,\textrm{B}}} }  \\
   {\rho_B \sqrt{ N_{a,\textrm{B}} N_{e,\textrm{B}} } } & N_{e,\textrm{B}}   \\
\end{array}} \right]. \nonumber
\end{eqnarray} 
The combining weights for $L_{e, \textrm{B}}(x)$ and $L_{a, \textrm{B}}(x)$ in (\ref{eq:apost_branch2}) are derived from the eigenvectors of $\mathbf{R}$. 
Write $\mathbf{R}$ as
$\mathbf{R} = \mathbf{G} \mathbf{\Lambda} \mathbf{G}^{-1}$ where $\mathbf{G}$ is a unitary matrix consisting of the eigenvectors of $\mathbf{R}$ and $\mathbf{\Lambda}$ is a diagonal matrix whose elements are the eigenvalues of $\mathbf{R}$. It is well-known that the transformation $[Y'_{e,\textrm{B}} \, \, \, Y'_{a,\textrm{B}}]^T = \mathbf{G}^{T} [Y_{e,\textrm{B}} \, \, \, Y_{a,\textrm{B}}]^T$ yields outputs whose correlation matrix is given by $\mathbf{\Lambda}$. 

Introducing a new variable $\lambda \triangleq \sqrt{N_{a,\textrm{B}} / N_{e,\textrm{B}}}$, (\ref{eq:apost_branch3}) can be rewritten as
\begin{eqnarray}
L_{a, \textrm{M}}(x) & = & \left( \dfrac{  1 - \rho_B \lambda^{-1}  } {1-\rho_B^2  } \right) L_{e,\textrm{B}}(x) + \left( \dfrac{   \rho_B^2 - \rho_B \lambda  } {1-\rho_B^2  } \right) L_{a,\textrm{B}}(x) . \label{eq:original_Lam}
\end{eqnarray} 

This equation allows us to construct the \textit{a priori} LLR for the main equalizer based on two correlated LLRs $L_{e, \textrm{B}}(x)$ and $L_{a, \textrm{B}}(x)$, along with the values of $\lambda$ and $\rho_B$ which can be estimated easily.
In our simulation, however, we observe that occasionally $L_{e, \textrm{B}}(x_n)$ and $L_{a, \textrm{B}}(x_n)$ take opposite polarities, due to large local noise values. When this happens, the polarity of $L_{a, \textrm{M}}(x_n)$ may also be inconsistent with that of $L_{e, \textrm{B}}(x_n)$, depending on the particular values of $\lambda$ and $\rho_B$ as well as the magnitudes of $L_{e, \textrm{B}}(x_n)$ and $L_{a, \textrm{B}}(x_n)$. It turns out that these events degrade the overall performance significantly.
Luckily, we find that a small modification handles this issue effectively and provides robust performance.
The modification is based on imposing a constraint that the polarities of $L_{a, \textrm{M}}(x_n)$ and $L_{e, \textrm{B}}(x_n)$ remain the same for all $n$ while following the rule set forth by (\ref{eq:original_Lam}) as much as possible. How do we achieve this?
A trick is to force a linear relationship, $\hat{L}_{a, \textrm{M}}(x) =  \alpha L_{e,\textrm{B}}(x)$ where $\alpha$ is a positive scaling factor that in general depends on $\rho_B$, 
and then find the value of $\alpha$ that would minimize the mean-squared error (MSE) between $L_{a,\textrm{M}}(x)$ as given in (\ref{eq:original_Lam}) and its forced approximation $\hat{L}_{a, \textrm{M}}(x)$. 

Denoting the MSE by $J$, we can write
\begin{eqnarray}
J & = & \textrm{E} \left[  \left| \left ( \dfrac{  1 - \rho_B \lambda^{-1}  } { 1-\rho_B^2  } - \alpha \right ) L_{e,\textrm{B}}(x)  + \left( \dfrac{   \rho_B^2 - \rho_B \lambda  } { 1-\rho_B^2  } \right) L_{a,\textrm{B}}(x) \right |^2 \right] \nonumber \\
& = & \textrm{E} \left[  \left | \left ( \dfrac{  1 - \rho_B \lambda^{-1} } { 1-\rho_B^2 } - \alpha \right ) \dfrac{2 Y_{e, \textrm{B}}} {N_{e, \textrm{B}}}  + \left( \dfrac{   \rho_B^2 - \rho_B \lambda  } { 1-\rho_B^2  } \right) \dfrac{2Y_{a, \textrm{B}}} {N_{a, \textrm{B}}} \right |^2 \right] \nonumber \\
& = & \dfrac{4}{N_{e, \textrm{B}}^2 N_{a, \textrm{B}}^2 (1-\rho_B^2 )^2 \lambda^2} \cdot \nonumber \\ 
& & \quad \Big[ \alpha^2  \left \{  \lambda^2 \left(1-\rho_B^2 \right)^2 N_{a, \textrm{B}}^2 (1+ N_{e, \textrm{B}})     \right \} \nonumber \\
& & \qquad - 2 \alpha \left \{ \lambda \left(1-\rho_B^2 \right) \left( \lambda - \rho_B  \right)  \left( N_{a, \textrm{B}}^2 (1+ N_{e, \textrm{B}})  - \lambda \rho_B N_{a, \textrm{B}} N_{e, \textrm{B}} (1+ \rho_B \sqrt{ N_{a, \textrm{B}} N_{e, \textrm{B}} } )  \right) \right \} \nonumber \\
& & \qquad + \left( \lambda - \rho_B  \right)^2 \Big( \lambda^2 \rho_B^2  N_{e, \textrm{B}}^2  (1+ N_{a, \textrm{B}}) + N_{a, \textrm{B}}^2 (1+ N_{e, \textrm{B}}) \nonumber \\ 
& & \qquad \quad -  2 \lambda \rho_B N_{a, \textrm{B}} N_{e, \textrm{B}} (1+ \rho_B \sqrt{ N_{a, \textrm{B}} N_{e, \textrm{B}} } ) \Big)   \Big] \nonumber
\end{eqnarray} 
where the second equality is due to the fact that for Gaussian variables, the log-likelihood function
can be expressed simply as a scaled version of the observation variable.  
Evaluating the derivative 
\begin{eqnarray}
\dfrac{\partial J}{\partial \alpha} & = & \dfrac{4}{N_{e, \textrm{B}}^2 N_{a, \textrm{B}}^2 (1-\rho_B^2 )^2 \lambda^2} \cdot \nonumber \\ 
& & \quad \biggr[ 2 \alpha  \left \{  \lambda^2 \left(1-\rho_B^2 \right)^2 N_{a, \textrm{B}}^2 (1+ N_{e, \textrm{B}})     \right \} \nonumber \\
& & \qquad - 2 \left \{ \lambda \left(1-\rho_B^2 \right) \left( \lambda - \rho_B  \right)  \left( N_{a, \textrm{B}}^2 (1+ N_{e, \textrm{B}})  - \lambda \rho_B N_{a, \textrm{B}} N_{e, \textrm{B}} (1+ \rho_B \sqrt{ N_{a, \textrm{B}} N_{e, \textrm{B}} } )  \right) \right \} \biggr].\nonumber \\
\end{eqnarray} 
and setting it equal to zero, we obtain the desired $\alpha$: 
\begin{eqnarray}
\alpha & = & \dfrac{\left( \lambda - \rho_B  \right)  \left\{ N_{a, \textrm{B}} (1+ N_{e, \textrm{B}})  - \lambda \rho_B N_{e, \textrm{B}} (1+ \rho_B \sqrt{ N_{a, \textrm{B}} N_{e, \textrm{B}} } )  \right\} } { \lambda \left(1-\rho_B^2 \right) N_{a, \textrm{B}} (1+ N_{e, \textrm{B}}) } .\label{eq:original_alpha}
\end{eqnarray} 

In \cite{Jeong10}, when the outputs of the forward and reversed DFEs were combined, it was shown that the sensitivity of the combiner output to the error in estimating the LLR correlation coefficient was reduced greatly if the variances of the two DFE outputs were assumed equal. Here we adopt the same strategy and assume that $N_{a,\textrm{B}} = N_{e,\textrm{B}} = N$ or $\lambda = 1$ to reduce the sensitivity of the combiner output to the error in estimating $\rho_B$. This assumption also reduces complexity, as $\lambda$ and $N$ no longer need be estimated. 
Accordingly, (\ref{eq:original_alpha}) becomes
\begin{eqnarray}
\alpha & = & \dfrac{  1+ N  - \rho_B (1+ \rho_B N )  } {  \left(1 + \rho_B \right) (1+ N) } \nonumber \\
& = & \dfrac{1- \rho_B}{1+ \rho_B} + \dfrac{ \rho_B (1- \rho_B) N   } {  \left(1 + \rho_B \right) (1+ N) } \label{eq:optimal_alpha}  \\
& \simeq & \dfrac{1- \rho_B}{1+ \rho_B}. \label{eq:approx_alpha}
\end{eqnarray} 
While the last approximation is valid when the noise correlation coefficient, $\rho_B$, is high or low and/or the noise variance is small, $N \ll 1$, 
empirical results indicate that (\ref{eq:approx_alpha}) gives robust performance at all situations.

In summary, we set the \textit{a priori} LLR for the main equalizer as a proper correlation-dependent scaling of the branch equalizer output LLR: 
\begin{eqnarray}
L_{a, \textrm{M}}(x) & = & \left ( \dfrac{1 - \rho_B} {1+ \rho_B} \right) L_{e,\textrm{B}}(x) . \label{eq:apriori_main_final}
\end{eqnarray}
Notice that here $\rho_B=0$ yields $L_{a, \textrm{M} }(x) = L_{e,\textrm{B}}(x)$ whereas $\rho_B=1$ leads to  
$L_{a, \textrm{M} }(x) = 0$. This is intuitively pleasing since when
the extrinsic LLR of the branch equalizer is completely uncorrelated with the \textit{a priori} LLR applied at the input of the branch equalizer,
the former should simply be passed on as the \textit{a priori} LLR of the main equalizer. On the other hand, a complete correlation would indicate that 
the extrinsic information out of the branch equalizer brings no new information to the main equalizer and the \textit{a priori} LLRs should all be set to zero.    

Equation (\ref{eq:apriori_main_final}) is also
valid in the opposite direction, i.e., when the \textit{a priori} LLR
of the branch equalizer is constructed from the extrinsic LLR of the main equalizer. That is, (\ref{eq:apriori_main_final}) is valid with the subscripts ``B" and ``M" swapped.

\subsection{Estimation of Noise Correlation Coefficient}
Assuming that the noise is stationary, the correlation
coefficient between $u_{a,\textrm{B}}$ and $u_{e,\textrm{B}}$ (or
$L_{a, \textrm{B} }(x)$ and $L_{e, \textrm{B} }(x)$) can be estimated through time-averaging $L_{a,
\textrm{B} }(x)$ and $L_{e, \textrm{B} }(x)$ over some reasonably
large finite window:
\begin{eqnarray}
\hat{\rho}_B & = & \dfrac{\sum_n \left \{  \big( L_{a, \textrm{B} }(x_n) - \textrm{sign}\left[ L_{a, \textrm{B} }(x_n)\right] m_{a,\textrm{B}} \big) \big( L_{e, \textrm{B} }(x_n) - \textrm{sign}\left[ L_{e, \textrm{B} }(x_n)\right] m_{e,\textrm{B}} \big) \right \} }{\sqrt{\sum_n \big( L_{a, \textrm{B} }(x_n) - \textrm{sign}\left[ L_{a, \textrm{B} }(x_n)\right] m_{a,\textrm{B}} \big)^2} \sqrt{\sum_n \big( L_{e, \textrm{B} }(x_n) - \textrm{sign}\left[ L_{e, \textrm{B} }(x_n)\right] m_{e,\textrm{B}} \big)^2}} \label{eq:rho}
\end{eqnarray}
where $m_{a,\textrm{B}} = \textrm{E}(L_{a, \textrm{B} }(x_n) | x_n
=+1)$ and $m_{e,\textrm{B}} = \textrm{E}(L_{e, \textrm{B} }(x_n) |
x_n =+1)$; due to symmetry it is also assumed that $\textrm{E}(L_{a, \textrm{B} }(x_n) | x_n=+1)=-\textrm{E}(L_{a, \textrm{B} }(x_n) | x_n=-1)$ and $\textrm{E}(L_{e, \textrm{B} }(x_n) |x_n =+1)=-\textrm{E}(L_{e, \textrm{B} }(x_n) |x_n =-1)$.

We can estimate the conditional means through time-averaging:
\begin{eqnarray}
\hat{m}_{a,\textrm{B}} & = & \frac{1}{2} \left\{ \overline{ \overline{ \left (  L_{a, \textrm{B} }(x_n) | L_{a, \textrm{B} }(x_n) \geq 0 \right)}} - \overline{\overline{ \left (  L_{a, \textrm{B} }(x_n) | L_{a, \textrm{B} }(x_n) < 0 \right)}} \right\}\nonumber \\
\hat{m}_{e,\textrm{B}} & = & \frac{1}{2} \left\{ \overline{\overline{ \left (  L_{e, \textrm{B} }(x_n) | L_{e, \textrm{B} }(x_n) \geq 0 \right)}} - \overline{\overline{ \left (  L_{e, \textrm{B} }(x_n) | L_{e, \textrm{B} }(x_n) < 0 \right)}} \right\}\nonumber
\end{eqnarray}
where $\overline{ \overline{u} }$ denotes the time-average of $u$. Note that
the signs of the LLRs between the main
equalizer output and the branch equalizer output might be different; in
estimating the correlation coefficient, we only consider those LLR
samples for which $\textrm{sign}\left[ L_{a, \textrm{B}
}(x_n)\right]$ and $\textrm{sign}\left[ L_{e, \textrm{B}
}(x_n)\right]$ are identical.

\subsection{Extension to the Case of Multiple Branch Equalizers}
Let us assume that there are one main equalizer and two branch
equalizers. Then, the \textit{a priori} LLR to the main equalizer
from each branch equalizer can be formulated as
\begin{eqnarray}
L^{(1)}_{a, \textrm{M} }(x)  =  \left( \dfrac{1 - \rho_B^{(1)}} {1+ \rho_B^{(1)}} \right) L^{(1)}_{e,\textrm{B}}(x) , \qquad
L^{(2)}_{a, \textrm{M} }(x)  =  \left( \dfrac{1 - \rho_B^{(2)}} {1+ \rho_B^{(2)}} \right) L^{(2)}_{e,\textrm{B}}(x) . \nonumber
\end{eqnarray}
where superscript $(i)$ points to a specific branch equalizer and
$\rho_B^{(i)}$ is the correlation coefficient between
$L^{(i)}_{e,\textrm{B}}(x)$ and $L^{(i)}_{a,\textrm{B}}(x)$.
Again, since the \textit{a priori} information (or extrinsic
information) can be modeled as an equivalent AWGN channel output,
under the assumption that the noise variances are the same, the
whitened/combined \textit{a priori} information to the main
equalizer can be shown to be
\begin{eqnarray}
L_{a, \textrm{M} }(x) & = &  \dfrac{1} {1+ \xi} \Big(  L^{(1)}_{a, \textrm{M} }(x) +  L^{(2)}_{a, \textrm{M} }(x) \Big) \label{apriori_main}
\end{eqnarray}
where $\xi$ is the noise correlation coefficient between
$L^{(1)}_{a, \textrm{M} }(x)$ and $L^{(2)}_{a, \textrm{M} }(x)$, which 
can be also estimated through time-averaging based on an
equation similar to (\ref{eq:rho}). It is straightforward to
extend the SISE algorithm to a system consisting of $\mu$ branch
equalizers via noise-whitening transformation.

\subsection{SISE Algorithm}\label{SISE}
Finally, the proposed SISE algorithm for an uncoded system can be
summarized as follows:
\begin{itemize}
\item Initialize the \textit{a priori} information of the main equalizer, i.e.,
$L_{a,\textrm{M}}(x_n)=0$ and $L^{(i)}_{a,\textrm{M}}(x_n) = 0$
for all time index $n$ and branch index $i$.
\item For the specified number of self-iterations,
\begin{enumerate}
\item Generate the extrinsic information of the main equalizer, $L_{e,\textrm{M}}(x_n)$, with the \textit{a priori} information $L_{a,\textrm{M}}(x_k)$ for all $k \neq n$ (the process of generating the extrinsic information in a given equalizer is described in Section V).
\item Compute the noise correlation coefficients, $\rho^{(i)}_{\textrm{M}}$, between $L_{e,\textrm{M}}(x_n)$ and $ \alpha_i L^{(i)}_{a,\textrm{M}}(x_n)$ where $\alpha_i$ is the combiner weight used on $i^{\rm{th}}$ branch equalizer output in constructing $L_{a,\textrm{M}}(x_n)$ in the previous cycle, $L_{a,\textrm{M}}(x_n) = \sum_i \alpha_i L^{(i)}_{a,\textrm{M}}(x_n)$, and set $L^{(i)}_{a,\textrm{B}}(x_n) = (1-\rho^{(i)}_{\textrm{M}}) /(1+\rho^{(i)}_{\textrm{M}}) \cdot L_{e,\textrm{M}}(x_n)$ for all $i$.
\item Generate the extrinsic information of each branch equalizer, $L^{(i)}_{e,\textrm{B}}(x_n)$, with the given \textit{a priori} information $L^{(i)}_{a,\textrm{B}}(x_k)$ for all $k \neq n$, for all $i$.
\item Compute the noise correlation coefficient, $\rho^{(i)}_{\textrm{B}}$, between $L^{(i)}_{e,\textrm{B}}(x_n)$ and $L^{(i)}_{a,\textrm{B}}(x_n)$ and set $L^{(i)}_{a,\textrm{M}}(x_n) = (1-\rho^{(i)}_{\textrm{B}})/(1+\rho^{(i)}_{\textrm{B}}) \cdot L^{(i)}_{e,\textrm{B}}(x_n)$ for all $i$.
\item Generate the \textit{a priori} information for the main equalizer, $L_{a,\textrm{M}}(x_n)$, from $L^{(i)}_{a,\textrm{M}}(x_n)$ via the extended equation of (\ref{apriori_main}).
\end{enumerate}
\end{itemize}

\section{Application to Turbo Equalization}\label{sec:iterSISE}
In this section, we propose turbo equalization based on the
previously developed SISE algorithm. Various iterative equalization algorithms are
possible based on this structure, but two main algorithms are
considered here.

\subsection{SISE 1 and SISE 2 Algorithms}\label{sec:iterSISE1}
\begin{figure}[!t]
\centering 
\subfigure[] 
{
\includegraphics[width=8.0cm]{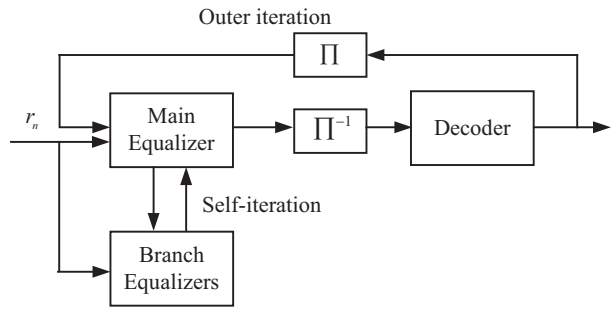}\label{fig:SISE1_info_flow}
}
\subfigure[] 
{
\includegraphics[width=8.0cm]{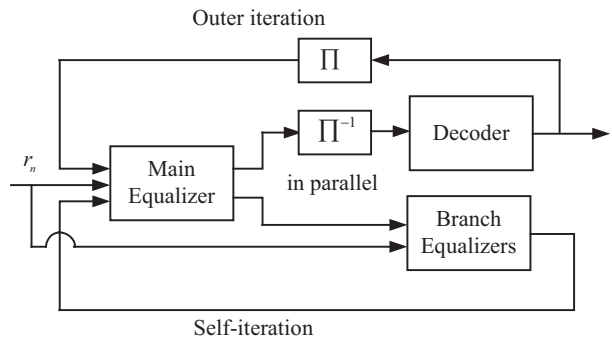}\label{fig:SISE2_info_flow}
} \caption{Information flow: (a) Turbo SISE 1 (b) Turbo SISE 2.}
\end{figure}

The first algorithm passes the uncorrelated extrinsic information
of the main equalizer to the branch equalizers and then to the
decoder in turn. The information flow of the SISE 1 algorithm is shown in Fig. \ref{fig:SISE1_info_flow}.
While the time sequence is not clear in the figure, we note that 
the main equalizer passes the soft information to the decoder \textit{after} one self-iteration is performed with its branch equalizer,  

Due to the sequential nature of the self-iteration steps, the SISE 1 algorithm has
a long latency issue; the second algorithm
(SISE 2) is also proposed to get around this issue. In contrast to
the first algorithm, SISE 2 passes the correlation-compensated extrinsic information of the main
equalizer to the branch equalizers and to the decoder
\textit{simultaneously}. Thus, in this case, the self-iteration step is performed in parallel with the outer turbo iteration. The information flow of SISE 2 is shown in Fig. \ref{fig:SISE2_info_flow}.

\subsection{Comparison of Complexity and Latency}
Let the computational complexity of the main equalizer, the branch
equalizers, and the decoder be $C_M$, $C_B$, and $C_D$,
respectively. For each outer iteration performed, the amount of
computation for the conventional turbo equalization is $C_M+C_D$,
whereas it is $2C_M+C_B+C_D$ and $C_M+C_B+C_D$ for SISE 1 and SISE
2, respectively. Moreover, assuming the processing time for the
main equalizer, the parallel branch equalizers, and the decoder is
all equal to $T$, the total processing time for each outer
iteration is $2T$, $4T$, and $2T$ for the conventional system,
SISE 1, and SISE 2, respectively. As will be shown later, in
addition to having complexity/latency advantage, SISE 2 also has
performance advantage over SISE 1 in many channel situations and
thus seems to be the preferred choice. Also, while SISE 2 requires
higher complexity (by $C_B$) than existing turbo equalizers when
$C_M$ is fixed in both cases, it will be shown that SISE 2 often
enables substantial error rate reduction in channel conditions
where the existing turbo equalizer cannot provide any performance
improvement regardless of how large $C_M$ is allowed to grow.

\section{Individual Constituent Equalizers for SISE}\label{sec:Filters}
Each constituent equalizer generates its own extrinsic LLR sequence given 
the estimated \textit{a priori} LLR sequence as well as the channel observation sequence
available at its input.
While any equalizer can be used as a constituent equalizer of the SISE,
we assume that the LE, the DFE or the BiDFE plays the role of each constituent equalizer in this paper.
For practical reasons, we also assume that a constituent equalizer can be based on either a time-invariant (TI) filter or a quasi-time-invariant (QTI) filter with tap weight setting changing once per outer turbo iteration or self-iteration.
As shown by the authors of \cite{TE02, TSK02}, the classical MMSE equalizer design can be modified by incorporating the mean and variance of the channel input symbols estimated via the extrinsic symbol information generated by the decoder. 
In Section \ref{sec:SISO equalizers}, we provide brief overviews on these equalizers, and in Section \ref{sec:Performance}, we discuss performance potentials of QTI versus TI filter structures at the limit of a large number of iterations.
  
\subsection{MMSE Filters Utilizing \text{A Priori} Information} \label{sec:SISO equalizers} 

\subsubsection{Linear Equalizer of \rm{\cite{TE02}}}
Here we basically summarize the LE approach of \cite{TE02}. In the process, however, we clarify specific parameter settings for the simulation results presented in this paper. Let the filter coefficient vector be 
${\bf{c}} = [c_{L_c }  \cdots c_0  \cdots c_{ - L_f } ]^T$ and define the observation vector
$ {\bf{r}} \buildrel \Delta \over = [r_{n - L_c }  \cdots r_n  \cdots r_{n + L_f } ]^T $. The filter output at time $n$, $y_n={\bf{r}}^T {\bf{c}}$, represents an estimate for the input symbol $x_n$. The corresponding estimation error $e_n  = x_n  - {\bf{r}}^T {\bf{c}}$, however, may have a non-zero mean if the average symbol value is non-zero, which is the case when the \textit{a priori} probabilities exist for the input symbols. 
To force the mean of the equalizer error to zero, the LE output is modified to  
$ y_n=({\bf{r}} - {\bf{\bar r}})^T {\bf{c}} + \bar x_n $ so that the error is given by  
$ e_n  = x_n  - \bar x_n  - ({\bf{r}} - {\bf{\bar r}})^T {\bf{c}}$.
The overbar denotes the statistical mean and indicates a vector of means when placed over a vector. It is straightforward to find the filter weights that minimize the mean-squared-error. Assuming the input symbols are independent, the MSE-minimizing taps weights at time $n$ are
\begin{equation}
\mathbf{c}_{opt}  = z_n \left[ {{\mathbf{H}\mathbf{A}\mathbf{H}^{\, T}}  + N_0 {\bf{I}}} \right]^{ - 1} {\mathbf{h}} \label{eq:copt}
\end{equation}
where  $z_n$ is the input symbol variance: $z_n  \triangleq \overline { ( x_n  - \overline{x}_n  )^2 }  = \overline{x^2_n}  - [\overline {x}_n]^2  = 1 - [\overline {x}_n]^2$, $\mathbf{H}$ is the 
$ L \times (L + L_h ) $ channel response matrix with $L = 1 + L_c  + L_f$, 
\begin{eqnarray}
\bf{H}= \left[ {\begin{matrix}
   {h_{L_h } } &  \cdots  & {h_1 } & {h_0 } & 0 &  \cdots  & 0  \cr 
   0 & {h_{L_h } } &  \cdots  & {h_1 } & {h_0 } &  \cdots  & 0  \cr 
    \vdots  & {} & {\ddots} &  \vdots  & {} & {\ddots} &  \vdots   \cr 
   0 &  \cdots  & 0 & {h_{L_h } } &  \cdots  & {h_1 } & {h_0 }  \cr 
 \end{matrix} } \right], \label{eq:H} 
\end{eqnarray}
${\bf{A}} = {\rm{Diag}}\,[z_{n - L_c  - L_h}  \cdots z_n  \cdots z_{n + L_f}  ]$, and ${\bf{h}} =  \, {\rm{[}}\underbrace {0 \cdots 0}_{L_c \, {\rm{zeros}}} \, \, h_0 \cdots h_{L_f} ]^T 
$ is the $(1 + L_c )^{\rm{th}}$ column of $\mathbf{H}$ from the left. Clearly the weight vector of (\ref{eq:copt}) is time-varying. In the classical MMSE-LE solution, $x_n$ is assumed to be equally likely, meaning $\overline {x}_n=0$ or $z_n=1$ for all $n$. This would have reduced (\ref{eq:copt}) to a time-invariant solution. Now, to complete the solution useful for turbo equalization, recall from Section \ref{sec:System_Model} that $y_n$ is to be generated while suppressing the current \textit{a priori} LLR $L_a(x_n)$ sample to zero. Setting $L_a(x_n)=0$ is equivalent to setting $\overline{x}_n=0$, which in turns leads to $z_n=1$. The final MMSE filter solution then can be constructed by effectively replacing all $z_{n}$ terms appearing in (\ref{eq:copt}) by 1.

In an effort to reduce the complexity associated with time-varying filter implementation, however, the filter coefficients can be made to change only once per turbo iteration.
A possible approximation for this is to replace every $z_k$ in (\ref{eq:copt}) by the time average:
$\overline{\overline{z}} = 1/N \sum_{k=0}^{N-1} z_k$ for the given iteration stage ($N$ is the codeword size)  \cite{TSK02}. Let the corresponding QTI filter coefficients be ${\bf{c}}_{qti}$:
\begin{equation}
{\mathbf{c}}_{qti}  =  \left[ \mathbf{H}\mathbf{\overline{\overline{A}}}\mathbf{H}^{\,T}  + N_0 {\mathbf{I}} \right]^{ - 1} {\mathbf{h}} \label{eq:cqti}
\end{equation}
where $\mathbf{\overline{\overline{A}}}$ is the matrix replacing all $z_{n}$ terms appearing in $\bf{A}$ with $\overline{\overline{z}}$ except replacing $z_n$ corresponding to the current time index by 1. Clearly, ${\mathbf{c}}_{qti}$ is a time-invariant vector for a given value of $\overline{\overline{z}}$. The equalizer output is then 
\begin{eqnarray}
y_n=({\bf{r}} - {\bf{\bar r}})^T {\bf{c}}_{qti}  \label{eq:ynqti}
\end{eqnarray}
with the computation of $\overline {r}_i=\sum\nolimits_l h_l \overline{x}_{i - l}$ in vector ${\bf{\bar r}}$ still based on the time-varying mean $ \overline{x}_k = (1-\delta_{k-n})\tanh \left\{ L_a (x_k )/2 \right\}$, where $\delta_l=1$ for $l=0$ and $\delta_l=0$ for $l \neq 0$.

Assuming that $y_n$ so obtained can be approximated as Gaussian when conditioned on a given value of the binary input symbol, the extrinsic LLR can be computed from (\ref{eq:Le_xn}). Write the equalizer output as $y_n = \beta x_n + v_n$, where $\beta$ is a scaling factor that can be easily shown to be 
$\beta = {\bf{h}}^T {\bf{c}}_{qti}$ and $v_n$ is the combination of the random noise and residual ISI, which is approximated as zero-mean Gaussian with variance 
\begin{eqnarray}
\sigma _{v_n}^{\,2} & = & {\mathop{\rm var}} [({\bf{r}} - {\bf{\bar r}})^T {\bf{c}}_{qti} ]|_{z_n  = 0} 
 =  {\bf{c}}_{qti}^T \left[ {{\bf{HA'}}_{} {\bf{H}}_{}^{\,T}  + N_0 {\bf{I}}} \right] {\bf{c}}_{qti}  \label{eq:LE v var}
\end{eqnarray}
where ${\bf{A'}} = {\rm{Diag}}\,[z_{n - L_c  - L_h}  \cdots z_{n-1} \, \, 0 \, \, z_{n+1} \cdots z_{n + L_f}  ]$. The extrinsic LLR is set as $L_e(x_n) = 2 \beta y_n /\sigma _{v_n}^{\,2}$. Note that while $\beta$ is time-invariant,
$\sigma _{v_n}^{\,2}$ is time-varying and so is ${\bf{\bar r}}$ in the computation of $y_n$. While $\sigma _{v_n}^{\,2}$ can also be approximated by a quasi-time-invariant quantity (by replacing ${\bf{A'}}$ in (\ref{eq:LE v var}) with ${\rm{Diag}}\,[\, \overline{\overline{z}}  \cdots \overline{\overline{z}} \, \, 0 \, \, \overline{\overline{z}} \cdots \overline{\overline{z}} \,]$) \cite{TSK02}, our simulation results in this paper reflect the time-varying $\sigma _{v_n}^{\,2}$ of (\ref{eq:LE v var}).  

\subsubsection{Decision Feedback Equalizer of \rm{\cite{Jeong10}}} 
While MMSE-DFE design is well-established and not much could be added to the existing body of knowledge, it is worth clarifying the assumption made on the past decision statistics in the derivation of the forward and feedback filter taps in DFE. In \cite{TSK02}, DFE filter taps are obtained assuming that the variances associated with the past known symbols (due to perfect decisions) are all zero. Here we take a view that the statistics for the past decisions should be identical to the actual input symbol statistics, which is more in tune with the underlying principle of DFE design.
It is in fact not necessary to assume zero-variances for the past symbols to derive the same results. We will also briefly summarize the technique of \rm{\cite{Jeong10}} for suppressing error propagation. 

Let the forward and feedback filter coefficient vectors be $ \mathbf{c} = \left[ c_{0} \, \, c_{-1} \cdots c_{-L_f} \right]^T$ and $\mathbf{d} = \left[ d_{L_d} \, \, d_{L_d-1} \cdots  d_1 \right]^T$, respectively. Given correct past decisions, past channel output (observation) samples are not correlated with the current and future observation samples while having no dependency on the current input. Thus, past observation samples do not provide any useful information for making decision on the current input symbol. Accordingly no causal forward taps are necessary. Also defining the composite vectors  ${\bf{u}} \triangleq [x_{n - L_d } \cdots x_{n - 1} \, | \, r_n \cdots r_{n + L_f } ]^T  \triangleq [{\bf{x}}_{-}^{T} \, | \, {\bf{r}}^{T}]^T$ and 
${\bf{f}} \buildrel \Delta \over = [ -d_{L_d }  \cdots  -d_1 \, | \, c_0  \cdots c_{ - L_f } ]^T  \buildrel \Delta \over = [ - {\bf{d}}^{T} \, | \, {\bf{c}}^{T}]^T$, the ideal-feedback DFE output with a zero-mean output error can be expressed as $y_n=({\bf{u}} - {\bf{\bar u}})^T {\bf{f}} + \bar x_n$. The MMSE solution can be obtained by solving $[\overline {({\bf{u}} - {\bf{\bar u}})({\bf{u}} - {\bf{\bar u}})^T } ]{\bf{f}}_{opt}  = \overline {(x_n  - \bar x_n )({\bf{u}} - {\bf{\bar u}})}$ or 
\begin{eqnarray}
\left[\begin{matrix}
   \bf{A} & \bf{B}  \cr 
   {\bf{B}}^T  & \bf{D}  \cr 
\end{matrix} \right] \left[\begin{matrix}
    - {\bf{d}}_{opt}   \cr 
   {\bf{c}}_{opt}   \cr 
\end{matrix} \right] =  \left[\begin{matrix}
   \bf{0}  \cr  \nonumber
   z_n {\bf{h}}  \cr 
\end{matrix} \right] \nonumber
\end{eqnarray}
where ${\bf{A}} = \overline {({\bf{x}}_ -   - {\bf{\bar x}}_ -  )({\bf{x}}_ -   - {\bf{\bar x}}_ -  )_{}^T }$, ${\bf{B}} = \overline {({\bf{x}}_ -   - {\bf{\bar x}}_ -  )({\bf{r}} - {\bf{\bar r}})^T }$, ${\bf{D}} = \overline {({\bf{r}} - {\bf{\bar r}})({\bf{r}} - {\bf{\bar r}})^T }$ and ${\bf{h}} = [h_0  \cdots h_{L_f} ]^T$. It  is easy to show that the solutions are: ${\bf{d}}_{opt}  = {\bf{A}}^{ - 1} {\bf{Bc}}_{opt} 
$ and ${\bf{c}}_{opt}  = z_n ({\bf{D}} - {\bf{B}}^T {\bf{A}}^{ - 1} {\bf{B}})^{ - 1} {\bf{h}}$.

Set $L_d=L_h$ (which lead to the optimal solution with the least number of coefficient taps) and define ${\bf{A}}_ +   = \overline {({\bf{x}}_ +   - {\bf{\bar x}}_ +  )({\bf{x}}_ +   - {\bf{\bar x}}_ +  )_{}^T }$ with ${\bf{x}}_ +   = [x_n  \cdots x_{n + L_f } ]^T$. Making use of the relationship 
${\bf{r}} = {\bf{H}}[({\bf{x}_ -}   - {\bf{\bar x}_ -}  )^{T} |({\bf{x}_ +}   - {\bf{\bar x}_ +} )^{\,T} ]^T, $ where $\mathbf{H}$ denotes the $ L \times (L + L_d ) $ channel response matrix of the form in (\ref{eq:H}) with $L = 1 + L_f$, we obtain 
\begin{eqnarray}
{\bf{c}}_{opt}  & = & z_n \left[ {\bf{H}}_2^{\, } {\bf{A}_ +}  {\bf{H}}_2^{\,T}  + N_0 {\bf{I}} \right]^{ - 1} {\bf{h}} , \qquad 
{\bf{d}}_{opt} =  {\bf{H}}_1^{\,T} {\bf{c}}_{opt} \label{eq:DFE_d}
\end{eqnarray}
where the reduced channel matrices ${\bf{H}}_1$ and ${\bf{H}}_2$ 
take the first $L_d$ columns and the remaining $1+L_f$ columns of $\bf{H}$, respectively, and ${\bf{A}}_ +   = {\rm{Diag}}\,{\rm{[}}z_n  \cdots z_{n + L_f } ]$. Note that this derivation does not resort to the assumption $\mathbf{A=0}$ made in \cite{TSK02}. The resulting taps in (\ref{eq:DFE_d}) are time-varying, so like in the case of LE, the filter coefficients are further constrained to be time-invariant at each iteration stage. This means that every $z_i$ in $\bf{A}_{+}$ in the computation of (\ref{eq:DFE_d}) 
is replaced by the time-average $\overline{\overline{z}} = (1/N) \sum_{k=0}^{N-1} z_k$ taken anew at each iteration stage. Let the corresponding QTI filter coefficient vectors be ${\bf{c}}_{qti}$ and 
${\bf{d}}_{qti}$:
\begin{eqnarray}
{\bf{c}}_{qti}  & = & \left[ {\bf{H}}_2^{\, } {\bf{\overline{\overline{A}}}_ +} {\bf{H}}_2^{\,T}  + N_0 {\bf{I}} \right]^{ - 1} {\bf{h}}, \qquad  
{\bf{d}}_{qti}  =  {\bf{H}}_1^{\,T} {\bf{c}}_{qti}. \label{eq:DFE_dqti}
\end{eqnarray}
where $\mathbf{\overline{\overline{A}}_+}$ is the matrix replacing all $z_{n}$ terms appearing in $\bf{A_+}$ with $\overline{\overline{z}}$ except $z_n=1$.

The DFE output is 
\begin{eqnarray}
y_n  = ({\bf{r}} - {\bf{\bar r}})^T {\bf{c}}_{qti} - ({\bf{\hat x}}_ -)^T {\bf{d}}_{qti}
\end{eqnarray}
with the computation of $\overline {r}_i=\sum\nolimits_l h_l \overline{x}_{i - l}$ in $\bf{\bar r}$ still based on the time-varying mean $ \overline{x}_k = (1- u_{k-n})\tanh \left\{ L_a (x_k )/2 \right\}$, where $u_l=1$ for $l\leq0$ and $u_l=0$ for $l > 0$. The vector $\bf{\hat x}_{-}$ consists of hard decisions on $L_d$ past symbols.

Instead of making the usual assumption that the feedback decisions are correct and that $y_n$ is conditionally Gaussian, the DFE design in \cite{Jeong10} considers the possibility of incorrect decisions affecting $y_n$ in formulating the extrinsic LLR. Write the DFE output as $y_n = \beta x_n + i_n + v_n$, where $\beta _n  = {\bf{h}}^T {\bf{c}}_{qti}$ with ${\bf{h}} = \,{\rm{[}}h_{\rm{0}} {\rm{ }} \cdots h_{L_f } ]^T$, $i_n$ is due to incorrect past decisions and $v_n$ is the combination of the random noise and residual ISI, which is again approximated as Gaussian. In \cite{Jeong10}, the extrinsic LLR is constructed as a combination of two conditional extrinsic LLRs computed for two separate cases: one without decision errors and one with decisions errors. Specifically, we have 
\begin{eqnarray}
L_e(x_n) & = & \ln \Bigg\{ \dfrac{ \exp \left( L_e(x_n | i_n = 0 ) \right) \mathrm{Pr}(i_n = 0) } { 1 + \exp \left( L_e(x_n | i_n =0 ) \right) } + \dfrac{ \exp \left( L_e(x_n | i_n \neq 0 ) \right) \mathrm{Pr}(i_n \neq 0) } { 1 + \exp \left( L_e(x_n | i_n \neq 0 ) \right) } \Bigg \} \nonumber \\
& & - \ln \Bigg\{ \dfrac{ \mathrm{Pr}(i_n = 0) } { 1 + \exp \left( L_e(x_n | i_n =0 ) \right) } + \dfrac{  \mathrm{Pr}(i_n \neq 0) } { 1 + \exp \left( L_e(x_n | i_n \neq 0 ) \right) } \Bigg \}. 
\end{eqnarray}
 
As usual, $L_e(x_n | i_n =0 )  = 2 \beta y_n /\sigma _{v_n}^{\,2}$, but for the case where $i \neq 0$, $L_e(x_n | i_n \neq 0 ) = \ln \left\{ \sum_{j} \frac{ \exp \left( L_e(x_n | i^{(j)}_n) \right) \mathrm{Pr}(i^{(j)}_n) } { \left\{ 1 + \exp \left( L_e(x_n | i^{(j)}_n) \right) \right\} \mathrm{Pr}(i_n \neq 0 ) } \right\} - \ln \left\{ \sum_{j} \frac{ \mathrm{Pr}(i^{(j)}_n ) } { \left\{ 1 + \exp \left( L_e(x_n | i^{(j)}_n) \right) \right\} \mathrm{Pr}(i_n \neq 0 ) } \right\} \simeq 2\varphi_n / (1+|\varphi_n|) $ where $i^{(j)}_n$ is the possible non-zero value of $i_n$ corresponding to the $j^\mathrm{th}$ error-pattern associated with the $L_d$ past symbols and $\varphi_n = \beta ( y_n - \overline{ (i_n | i_n \neq 0) } ) /\sigma _{v_n}^{\,2}$ \cite{Jeong10}. The conditional mean $\overline{ (i_n | i_n \neq 0) } = \overline{i}_n / \mathrm{Pr}(i_n \neq 0)$ can be estimated using the \textit{a posteriori} LLRs associated with $L_d$ past decisions. The variance of $v_n$ is obtained as
\begin{eqnarray}
\sigma _{v_n}^{\,2} & = & {\mathop{\rm var}} [({\bf{u}} - {\bf{\bar u}})^T {\bf{f}}_{qti} ]|_{z_n  = 0} 
 =  {\bf{c}}_{qti}^T \left[ {\bf{H}}_2^{ } {\bf{A'}}_{+} {\bf{H}}_2^{\,T}  + N_0 {\bf{I}} \right] {\bf{c}}_{qti}
\end{eqnarray}
where ${\bf{f}}_{qti} \buildrel \Delta \over = [ - {\bf{d}}_{qti}^{T} \, | \, {\bf{c}}_{qti}^{T}]^T$ and 
${\bf{A'}}_+  = {\rm{Diag}}\, [0  \, \, z_{n+1} \cdots z_{n+L_f} ]$. Again, $\sigma _{v_n}^{\,2}$ is time-varying here. 

\subsubsection{Bi-directional DFE of \rm{\cite{Jeong10}}}
The BiDFE of \cite{Jeong10} is based on using two DFEs running in opposite directions. The extrinsic LLR of each DFE is obtained using the method described above and combined to yield the BiDFE's extrinsic LLR:
\begin{eqnarray}
L_e(x_n) =  \dfrac{1} {(1+ \rho_{bi})} \Big( L_{e,f}(x_n) + L_{e,b}(x_n) \Big)
\end{eqnarray}
where the subscripts $f$ and $b$ mean the forward and backward DFEs, respectively, and $\rho_{bi}$ is the noise correlation coefficient between two DFEs and can be estimated using time-average, \rm{\cite{Jeong10}}.

\subsection{Performance Potentials of Suboptimal Equalizers Under Perfect \textit{A Priori} Information}\label{sec:Performance}
Let us consider the ideal condition for the equalizer where
the perfect \textit{a priori} information is available, i.e.,
$\overline{x}_n = x_n$ [or $ L_a(x_n) = \pm \infty$] for all $n$.
This condition may be satisfied when many iterations are
performed at sufficiently high channel SNRs in turbo equalization. 
Under this condition, it has already been discussed in \cite{TSK02} that the LE based on the time-varying optimal filter provides the ideal matched filter bound. Since the DFE and the BiDFE cannot be worse than the LE under this ideal condition, it follows that all equalizers with time-varying filters provide the matched filter performance.
It turns out that the same is true for the QTI-based equalizers as well, 
while the TI filters cannot achieve the matched filter performance even under the ideal condition.

Define the output SNR for the QTI-LE 
associated with an infinite number of turbo iterations as $\mathrm{SNR}_{\infty, \mathrm{QTI-LE}} = \lim_{\, \,\overline{\overline{z}} \, \to \,0} \beta^2/\sigma _{v_n}^{\,2}$,
assuming repeated iterations lead to perfect \textit{a priori} information (and thus perfect time-averaged \textit{a priori} information as well). Apparently, as $z_k \to 0$ for all $k$, $\overline{\overline{z}}\, \to \,0$. The filter weight vector of (\ref{eq:cqti}) then becomes ${\bf{c}}_{qti}  \, \to \, {\bf{h}}/\,(1+N_0)$ and the $v_n$ variance in (\ref{eq:LE v var}) takes the form $\sigma _{v_n}^{\,2} \to N_0  {\bf{c}}_{qti}^T {\bf{c}}_{qti}$. Since $\beta  = {\bf{h}}^T {\bf{c}}_{qti}$, it is easy to see that 
$\mathrm{SNR}_{\infty, \mathrm{QTI-LE}} = {\bf{h}}^T {\bf{h}}/{N_0} = 1/{N_0}  \label{SNR_QTI-LE}$ which corresponds to the matched filter performance. Again, the DFE gives the same result under the QTI constraint.
For the BiDFE, the noise
correlation coefficient between the normal DFE and the
time-reversed DFE is 1, which means they produce the same
equalized output and there is no SNR advantage of combining
\cite{Jeong10}. Accordingly, the LE, the DFE and the BiDFE all produce the same equalized outputs and the output SNRs with the QTI filters are given by
\begin{eqnarray}
\mathrm{SNR}_{\infty, \mathrm{QTI-LE}} = \mathrm{SNR}_{\infty, \mathrm{QTI-DFE}} = \mathrm{SNR}_{\infty, \mathrm{QTI-BiDFE}} = 1/N_0.
\end{eqnarray}
While all equalizer schemes can achieve the matched filter performance under the ideal condition, we stress that their realized performances are significantly different in practice \cite{Jeong10}. 

For the TI filters, on the other hand, the maximum achievable output SNRs
can be found as follows. First consider the LE; the filter weight vector is ${\bf{c}}_{tiLE}  = \left[ {\bf{HH}}_{}^{\,T}  + N_0 {\bf{I}} \right]^{ - 1} {\bf{h}}$. Keep in mind that the equalizers based on TI filters do not make use the \textit{a priori} information. As $z_k  \to 0$ for all $k$, $\sigma _{v_n}^{\,2} \to N_0  {\bf{c}}_{tiLE}^T {\bf{c}}_{tiLE}$. The output SNR thus becomes  
\begin{equation}
\mathrm{SNR}_{\infty, \mathrm{TI-LE}} = \mathop {\lim }\limits_{\overline{\overline{z}}\, \to \,0} \dfrac{\beta^2} {\sigma _{v_n}^{\,2}}  = \dfrac{1}{N_0} \dfrac{[{\bf{h}}^T {\bf{c}}_{tiLE}]^2} {{\bf{c}}_{tiLE}^T {\bf{c}}_{tiLE}}  = \dfrac{1}{N_0} \dfrac{[{\bf{h}}^T {\bf{P}} {\bf{h}}]^2}{{\bf{h}}^T {\bf{P}}^T{\bf{P}} {\bf{h}}} \label{SNR_TI-LE} \\
\end{equation}
where ${\bf{P}}=\left[ {\bf{HH}}_{}^{\,T}  + N_0 {\bf{I}} \right]^{ - 1}$. For the DFE, we get a similar expression:
\begin{equation}
\mathrm{SNR}_{\infty, \mathrm{TI-DFE}} = \dfrac{[{\bf{h}}^T {\bf{c}}_{tiDFE}]^2} {N_0 {\bf{c}}_{tiDFE}^T {\bf{c}}_{tiDFE}}  = \dfrac{1}{N_0} \dfrac{[{\bf{h}}^T {\bf{P}}_2 {\bf{h}}]^2}{{\bf{h}}^T {\bf{P}}_2^T{\bf{P}}_2^{\,} {\bf{h}}} \label{SNR_TI-DFE} \\
\end{equation}
where ${\bf{P}}_2=\left[ {\bf{H}}_{2}^{\,}{\bf{H}}_{2}^{\,T}  + N_0 {\bf{I}} \right]^{ - 1}$.
For the BiDFE, it can be shown that $\mathrm{SNR}_{\infty, \mathrm{TI-BiDFE}} = [2/(1 + \rho)] \mathrm{SNR}_{\infty, \mathrm{TI-DFE}}$ where $\rho$ is the noise correlation coefficient between
the normal DFE and the time-reversed DFE outputs under perfect \textit{a
priori} information and is given in \cite{Jeong10} as a fucntion of the two DFE filter coefficients. 
It is easy to see that
\begin{eqnarray}
\mathrm{SNR}_{\infty, \mathrm{TI-LE}} \leq \mathrm{SNR}_{\infty, \mathrm{TI-DFE}} \leq \mathrm{SNR}_{\infty, \mathrm{TI-BiDFE}}  \leq  1/N_0
\end{eqnarray}
where the equalities hold in the memoryless AWGN channel.

In order to incorporate the above output SNR analysis with the
EXIT chart analysis, one can compute the mutual information (MI) with these output
SNRs. Because only the Gaussian noise term remains in the
equalized output under the perfect \textit{a priori} information,
the MI is simply given as
\begin{eqnarray}
C_b (\mathrm{SNR}) & \triangleq & 1 - \int_{ - \infty }^{\infty}  \dfrac{e^{ - \tau^2 /2}} { \sqrt {2\pi } }  \log_2 \left \{ {1 + e^{ - 2\tau \sqrt{\mathrm{SNR}}   - 2\mathrm{SNR}} } \right\} d\tau   \label{eq:C(R)}
\end{eqnarray}
where $C_b(\mathrm{SNR})$ is the symmetric information rate of the
binary-input Gaussian channel for a given value of SNR. Note that the
MI computed by substituting $\mathrm{SNR} = \mathrm{SNR}_{\infty}$ in
(\ref{eq:C(R)}) is the maximum attainable MI by an equalizer. The
corresponding numerical results will be shown to be consistent with the 
the simulated MI trajectories
as will be presented for some selected cases toward the end of Section 
\ref{sec:Simulation Results}.

\section{Simulation Results and Discussion}\label{sec:Simulation Results}
In this section, simulation results of the proposed SISE
equalization schemes for both uncoded and coded systems are
presented. The transmitted symbols are modulated with binary
phase-shift keying (BPSK) so $x_n \in \{\pm 1\}$. The message bit
length is $2^{11}$. We also assume that the noise is AWGN, and the
noise variance and the channel impulse response are perfectly known to
the receiver. The channel we consider here are severe ISI channels that show deep nulls and valleys within the 
signal band. When the ISI is not severe, the LE scheme of \cite{TE02}
already provides near-optimal performance, as mentioned in the Introduction section.

\subsection{Uncoded System}
\begin{figure}[!t]
\centering
\includegraphics[width=14.0cm]{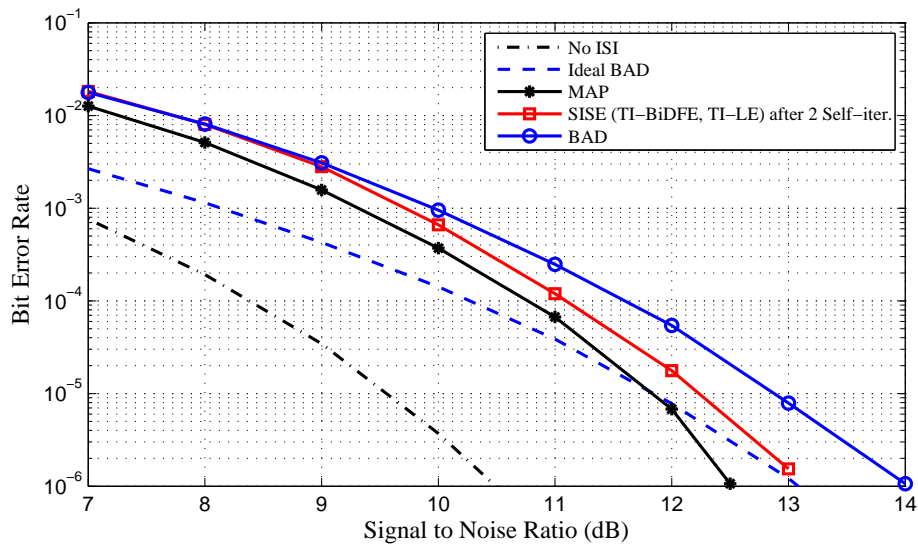}
\caption{BER curves on Channel $\mathbf{h_1}$.}\label{fig:BER1}
\end{figure}

The impulse response of the ISI channel $\mathbf{h_1}
=(1/\sqrt{6}) [1 \quad 2 \quad 1]^T$ discussed in \cite{BDC} is
used for the uncoded system. This channel with its system response (D-transform of the impulse response) given by $H_1(D)=(1/\sqrt{6})(1+D)^2$ has a second-order spectral null at the right edge of the Nyquist band, as shown in Fig. \ref{fig:Freq_channels}. 
Three different equalizer types are simulated for this channel. The SISE
method is the self-iterating soft equalizer algorithm described in
Section \ref{SISE}. Specifically, the SISO BiDFE algorithm of
\cite{Jeong10} is adopted as the main equalizer and the SISO LE of \cite{TE02} is used for the branch equalizer. 
Moreover, when the final hard decisions are released by the BiDFE,
the arbitration criterion of \cite{BAD05} with window size 15 is
employed; the symbol sequence is decided among the estimated
sequences of two DFEs based on which candidate shows the smaller
MSE in a window centered around the symbol of interest. Finally, 2 self-iterations are applied. 
The ``BAD" method is the algorithm of \cite{BAD05} employing two classical DFEs running in opposite directions with an arbitration strategy as described above.
The MAP equalizer is the optimal equalizer implemented via the well-known BCJR algorithm \cite{BCJR74}. 
Each DFE in the BiDFE or BAD consists of 13 feedforward taps and 2 feedback taps
($L_c = 12$ and $L_d = 2$) while the LE uses 15 taps ($L_c = 7$
and $L_f = 7$) for $\mathbf{h_1}$. Fig. \ref{fig:BER1} shows
the performance comparison. As the figure shows, the proposed SISE
algorithm shows the superior performance to the BAD method of
\cite{BAD05} and approaches the performance of ``Ideal BAD" with perfect past decisions in either direction.
Increasing the number of filter taps for BAD did not give any noticeable performance gain. 

\subsection{Coded System}
\begin{figure}[!t]
\centering 
\includegraphics[width=14cm]{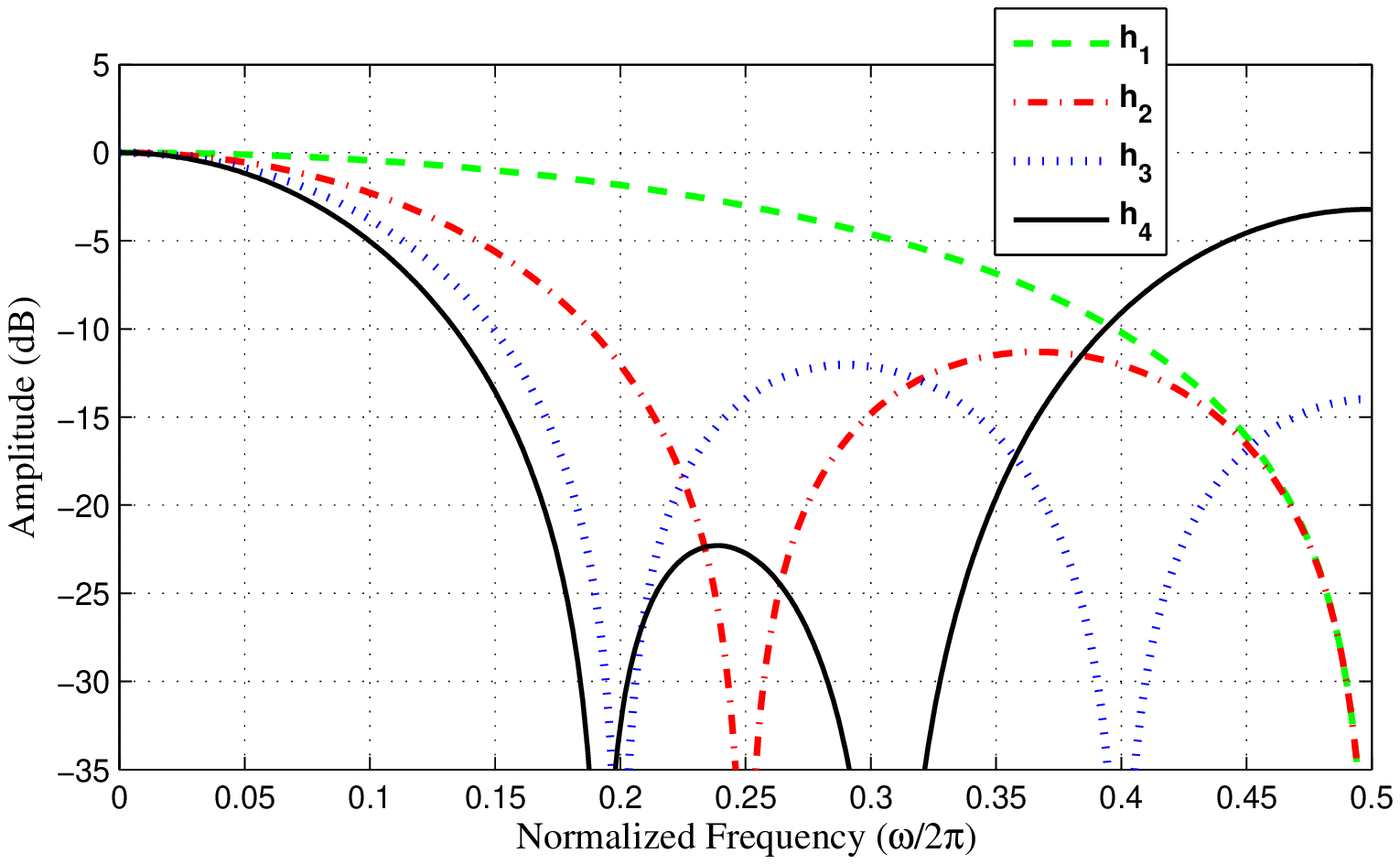}
\caption{Frequency Magnitude Response of the ISI Channels:
$\mathbf{h_1} =(1/\sqrt{6})
[1 \quad 2 \quad 1]^T$, $\mathbf{h_2}
=(1/\sqrt{44}) [1 \quad 2 \quad 3 \quad 4 \quad 3 \quad 2 \quad
1]^T$, $\mathbf{h_3}
=(1/\sqrt{85}) [1 \quad 2 \quad 3 \quad 4 \quad 5 \quad 4 \quad 3 \quad 2 \quad
1]^T$, and $\mathbf{h_4}
\approx 0.0795 [1 \quad 0.707 \quad 4 \quad 2.121 \quad 7.25 \quad 3 \quad 7.25 \quad 2.121 \quad 4 \quad 0.707 \quad 1]^T$.} 
 \label{fig:Freq_channels}
\end{figure}

In this subsection, simulation results of the iterative SISE
schemes are presented. The transmitted symbols are
encoded with a recursive rate-$1/2$ convolutional code encoder
with the parity generator polynomial $(1+D^2) / (1+D+D^2)$.
The size of the coded data packet (and thus the size of the interleaver used for turbo iteration between the equalizer and the decoder)
is $2^{12}$ bits. A random interleaver is employed. The
impulse response of the severe ISI channel $\mathbf{h_2}
=(1/\sqrt{44}) [1 \quad 2 \quad 3 \quad 4 \quad 3 \quad 2 \quad
1]^T$ investigated in \cite{Jeong10} as well as an even more sever
ISI channel of $\mathbf{h_3} =(1/\sqrt{85}) [1 \quad 2
\quad 3 \quad 4 \quad 5 \quad 4 \quad 3 \quad 2 \quad 1]^T$ are
used for evaluating and comparing the performances of iterative
equalizers. While both of these channels as well as $\mathbf{h_1}$ represent samples of the triangular time function, their spectral characteristics are quite different, as can be seen in Fig. \ref{fig:Freq_channels}. Their corresponding system polynomials are $H_2(D)= (1/\sqrt{44})(1+D)^2(1+D^2)^2$ and $H_3(D)= (1/\sqrt{85})(1+D+D^2+D^3+D^4)^2$, which reveal that these channels possess second-order nulls either within the Nyquist band or at its edge. We also select an additional channel whose time-domain shape is quite different from these channels.
Specifically, we place one second-order null and one third-order null well inside the Nyquist band according the system polynomial $H_4(D)
=(\sqrt{32/5061}) \left( 1+D/\sqrt{2}+D^2 \right)^3 \left( 1-D/\sqrt{2}+D^2 \right)^2$. Its magnitude response in frequency is also included in Fig. \ref{fig:Freq_channels}. The corresponding impulse response is $\mathbf{h_4}
=(\sqrt{32/5061}) [1 \quad 1/\sqrt{2} \quad 4 \quad 3/\sqrt{2} \quad 29/4 \quad 17/4\sqrt{2} \quad 29/4 \quad 3/\sqrt{2} \quad 4 \quad 1/\sqrt{2} \quad 1]^T$.

In the error rate figures, curves labeled by ``SISE 1" and ``SISE
2" correspond to the iterative SISE algorithms described in
Section \ref{sec:iterSISE}. The label ``(M, B)" in the legend denotes a
specific `M' algorithm as the main equalizer and a `B' algorithm
as the branch equalizer. For instance, ``SISE 2 (QTI-LE, QTI-DFE)"
denotes the iterative SISE 2 algorithm with the LE with
a quasi-time-invariant filter as the main equalizer and the
DFE with quasi-time-invariant filters as the
branch equalizer. The label ``TV-LE" denotes the MMSE linear equalizer with the
time-varying filter (or the ``exact MMSE" of \cite{TE02}). The label ``Ideal" indicates the
performance of an equalizer with perfect \textit{a priori}
information.

The straightforward LLR mapping method [$y_n$-to-$L_e(x_n)$ conversion] of \cite{TE02} is adopted
for the LE while the LLR mapping method of \cite{Jeong10} is
used for the DFE and BiDFE. Moreover, the DFE (and each DFE in the
BiDFE) consists of 21 feedforward taps and 6 feedback taps ($L_c =
20$ and $L_d = 6$) on $\mathbf{h_2}$; $L_c = 20$ and $L_d = 8$ on $\mathbf{h_3}$;
and $L_c = 20$ and $L_d = 10$ on $\mathbf{h_4}$. The
LE uses 27 taps ($L_c = 13$ and $L_f = 13$) for
$\mathbf{h_2}$, 29 taps ($L_c = 14$ and $L_f = 14$) for
$\mathbf{h_3}$, and 31 taps ($L_c = 15$ and $L_f = 15$) for
$\mathbf{h_4}$. Again, ``MAP" refers to the optimal equalizer
implemented via the BCJR algorithm. Finally, the decoder is implemented using the BCJR
algorithm and 20 turbo outer iterations are applied to achieve the full potential of each turbo equalization system.

\begin{figure}[!t]
\centering
\includegraphics[width=14.0cm]{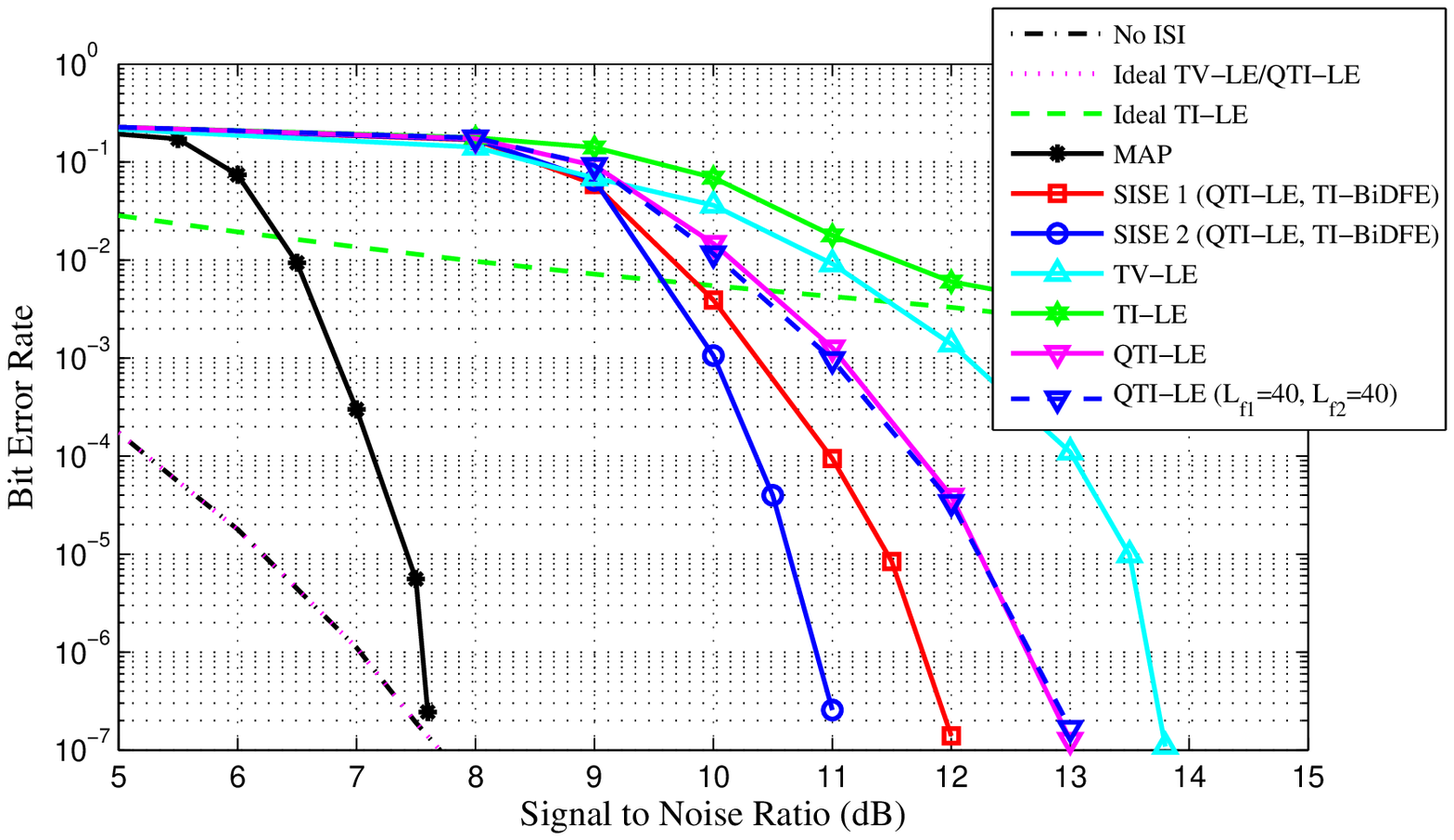}
\caption{LE-based BER curves on Channel $\mathbf{h_2}$ after 20 outer iterations.}\label{fig:BER2_LE}
\end{figure}

\begin{figure}[!t]
\centering
\includegraphics[width=14.0cm]{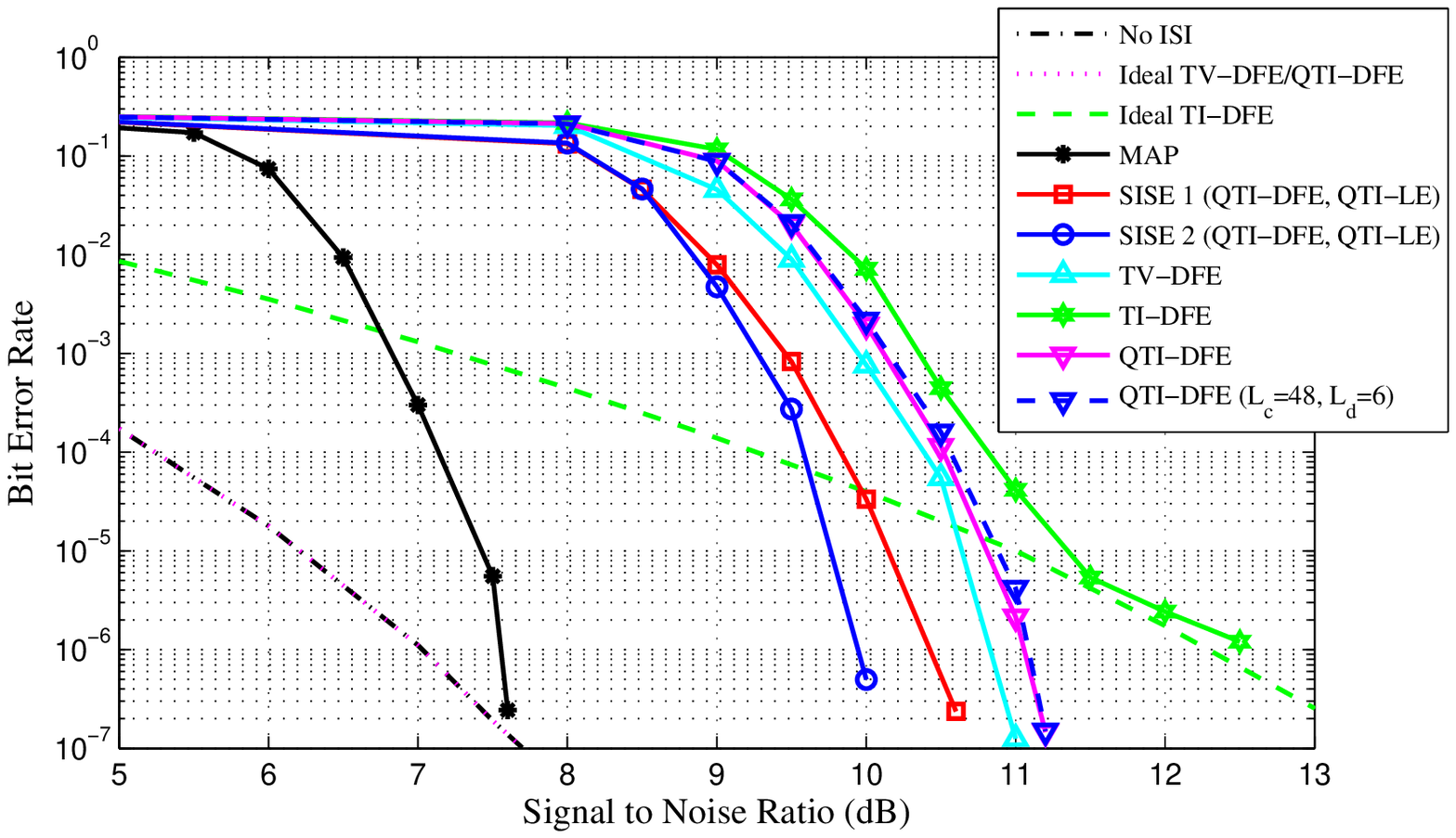}
\caption{DFE-based BER curves on Channel $\mathbf{h_2}$ after 20 outer iterations.}\label{fig:BER2_DFE_LE}
\end{figure}

\begin{figure}[!t]
\centering
\includegraphics[width=14.0cm]{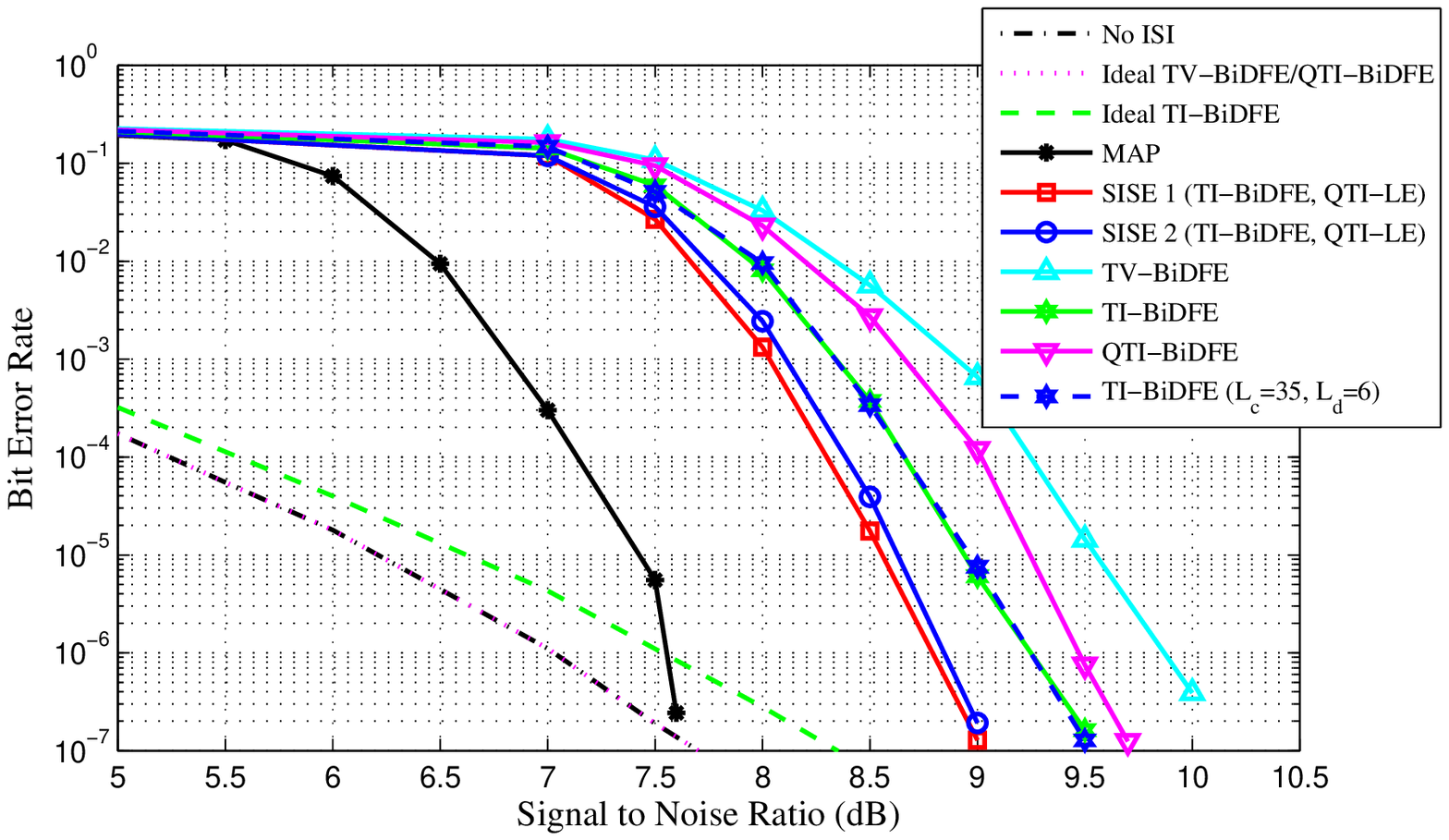}
\caption{BiDFE-based BER curves on Channel $\mathbf{h_2}$ after 20 outer iterations.}\label{fig:BER2_BiDFE}
\end{figure}

Figs. \ref{fig:BER2_LE}, \ref{fig:BER2_DFE_LE}, and
\ref{fig:BER2_BiDFE} show the performance of several turbo
equalizers on
$\mathbf{h_2}$. The performance of the proposed iterative SISE
algorithms is compared with the performance achieved when the
equalizer consists only of a single main equalizer.

Fig. \ref{fig:BER2_LE} shows performance comparison with LE-based
single equalizers. Among the single LE schemes, the QTI filter design
method gives the best performance, outperforming even the TV filter
method (the reason for which is explained below). When the number of filter taps increases to 81 (40 causal
and 40 anticausal taps) versus a total of 27, the QTI-LE method
does not provide any performance gain, indicating that using 27
taps for this channel already realizes the QTI-LE's full potential. Of
the two SISE methods showing superior performance to the single
LE, SISE 2 is better and is an obvious winner given its lower
complexity and latency compared to SISE 1. Both SISE schemes
employ TI-BiDFE as the sole branch equalizer. Given that each DFE
in BiDFE has a total of 27 taps,
each SISE scheme requires 81 taps overall.

Fig. \ref{fig:BER2_DFE_LE} tells a similar story but the comparison
is now with the DFE schemes. Accordingly, the main equalizer of the SISE methods is also set up with the DFE, under the QTI filter design scheme. The branch equalizer is a QTI-LE. 
Among the single equalizers, this time, the TV filter
method performs slightly better than the QTI filter scheme (again, increasing
the number of QTI filter taps does not give any performance boost).
Still, SISE methods give the best performance, other than the
MAP-based turbo equalizer, with SISE 2 once again coming ahead.  
Fig. \ref{fig:BER2_BiDFE} presents results that compare schemes
based on the BiDFE. The QTI-filter-based single equalizer is better than
the TV filter method and, again, increasing the filter length does not improve
performance. 
As for the branch equalizer, SISE schemes use the QTI-LE. 
This time, SISE 1 performs better than SISE 2, albeit by a
very small margin. 

A comparison of the simulation results in Figs. \ref{fig:BER2_LE}
and \ref{fig:BER2_BiDFE} reveals that the performance of ``SISE
(TI-BiDFE, QTI-LE)" is considerably better than that of ``SISE
(QTI-LE, TI-BiDFE)". On the other hand, among the single
equalizers, TI-BiDFE outperforms QTI-LE by a large margin. It turns
out that the extrinsic LLR quality of the main equalizer is an
important factor determining the overall performance of the SISE
algorithms, as the extrinsic LLRs of the main equalizer are passed
to the decoder as well as the branch equalizers. Therefore, in the
design of the SISE algorithm, the equalizer showing the best BER
performance (or LLR quality) should be chosen as the main
equalizer.

When different filter types are compared, the QTI filters
sometimes provide even better BER performance than the
TV filters (as shown in Figs. \ref{fig:BER2_LE} and \ref{fig:BER2_BiDFE}), which contradict the simulation results of \cite{TSK02}. 
This is mainly due to the fact that when inaccurate \textit{a priori} information arrives, the optimal
TV filters more easily fail to produce reliable extrinsic information than the QTI filters. 
In other words, the TV (or exact MMSE) filters of \cite{TSK02} are designed on the premise that the incoming \textit{a priori} information is accurate;
when the incoming \textit{a priori} information is not reliable as in severe ISI channels under consideration in this paper, they tend to generate low-quality LLRs. 
Furthermore, a few large incoming LLRs in wrong direction can degrade the overall turbo equalization performance quickly as we have observed during our simulation.
In the case of the QTI filter, a small number of large-magnitude LLRs tend to get averaged-down. Also, recall that an equalizer with QTI filters can also achieve the matched filter performance if enough turbo iterations are performed. Between the LE and the DFE, this effect is more pronounced with the LE, although even in the case of the DFE (as seen in Figs. \ref{fig:BER2_DFE_LE})
the performance advantage of TV design is small over the QTI method. In the case of the BiDFE, this sensitivity plays an even bigger role in
determining the overall performance. As indicated in Fig.
\ref{fig:BER2_BiDFE}, TI-BiDFE shows better performance than
BiDFE with any other filter types since it would be better not to
update the filter taps at all when the \textit{a priori}
information is unreliable at low SNRs.

\begin{figure}[!t]
\centering
\includegraphics[width=14.0cm]{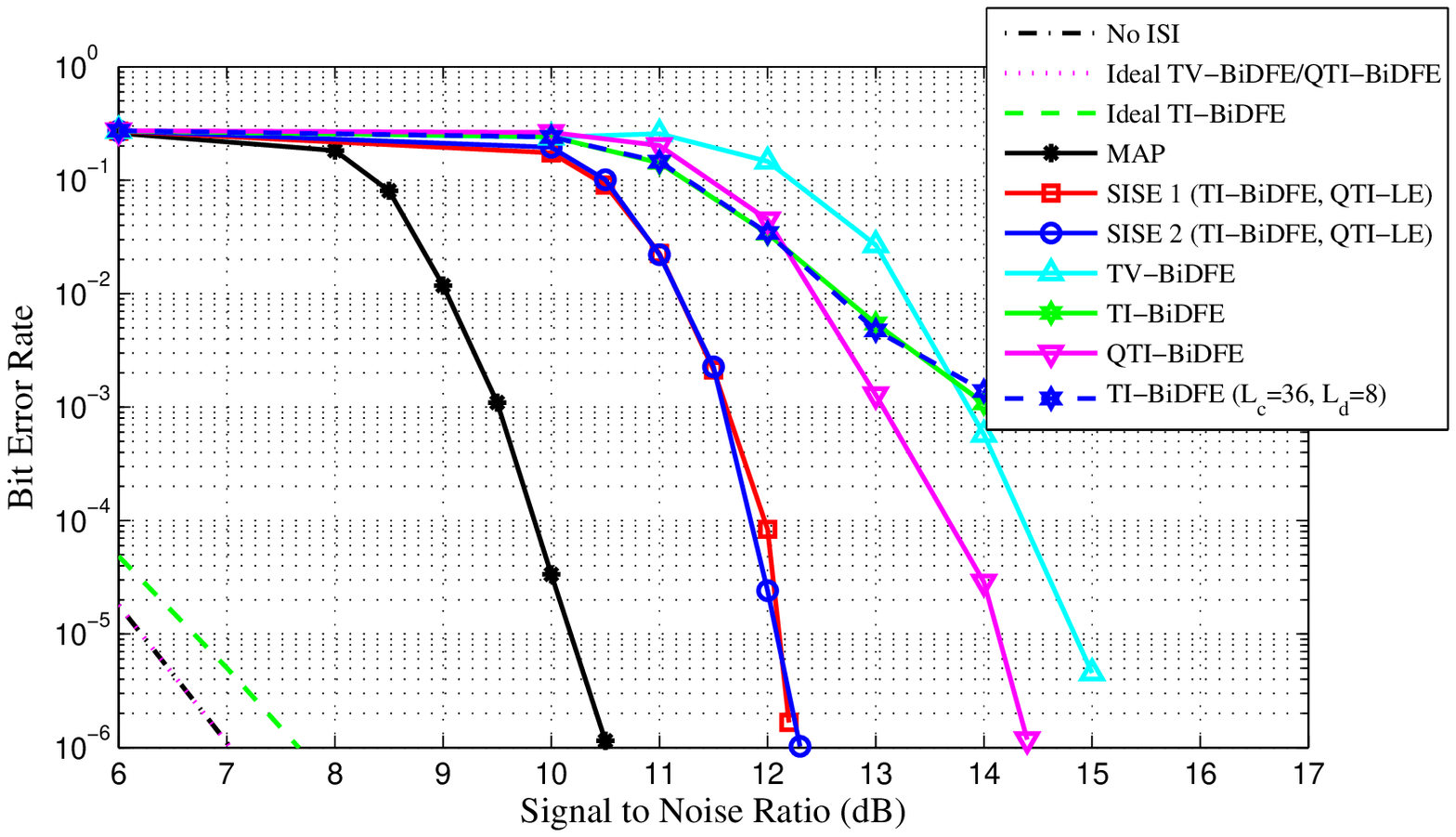}
\caption{BiDFE-based BER curves on Channel $\mathbf{h_3}$ after 20 outer iterations.}\label{fig:BER3_BiDFE}
\end{figure}

Fig. \ref{fig:BER3_BiDFE} shows comparison in the extremely severe
ISI channel of $\mathbf{h_3}$. Among the single equalizers, TI-BiDFE
schemes exhibit clear failure because the erroneously generated \textit{a priori} LLRs from the decoder during the iterations cause more errors in the subsequent turbo iterations, while QTI-BiDFE is better than
TV-BiDFE. Both SISE schemes show robust performance, again lagged
in performance only by the optimal MAP-based scheme.

The performance of the turbo equalizers are analyzed by using the
EXIT chart \cite{Brink01}, a diagram demonstrating the MI transfer
characteristics of the two constituent modules that exchange soft
information. In the EXIT charts, the behavior of the channel
equalizer is described with its input and output on the horizontal
and vertical axis, respectively, while the behaviour of the decoder
is described in the opposite way. The EXIT chart curves typically
define a path for the MI trajectory to move up during iterative
processing of soft information. Moreover, the number of stairs
that a given MI trajectory (averaged over 1000 sample codeword
blocks here) takes to reach the highest value indicates the
necessary number of iterations toward convergence.

\begin{figure}[!t]
\centering
\includegraphics[width=14.0cm]{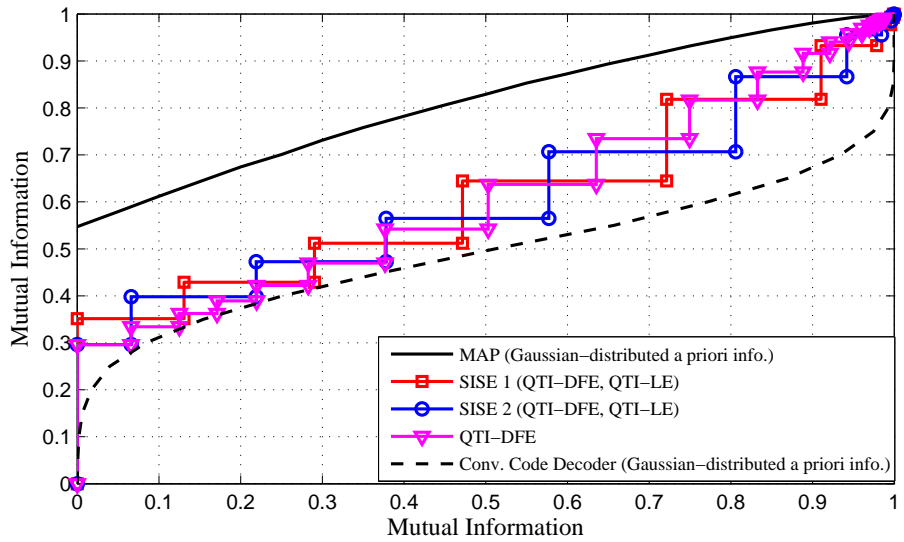}
\caption{DFE-based EXIT chart on Channel $\mathbf{h_2}$ at SNR=10 dB.}\label{fig:EXIT2_DFE_LE_10dB}
\end{figure}

\begin{figure}[!t]
\centering
\includegraphics[width=14.0cm]{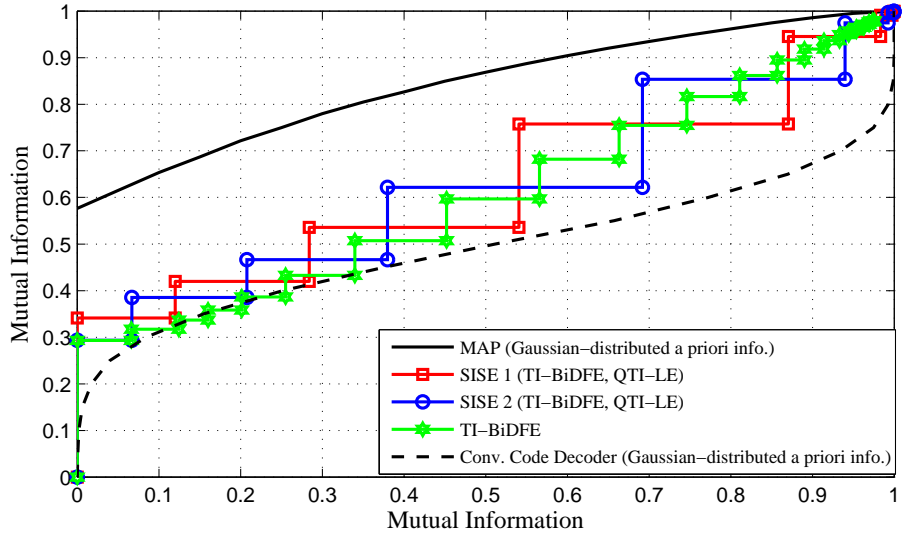}
\caption{BiDFE-based EXIT chart on Channel $\mathbf{h_3}$ at SNR=13 dB.}\label{fig:EXIT3_BiDFE_13dB}
\end{figure}

In order to avoid excessive cluttering, only the trajectories of
SISE 1, SISE 2 and the single equalizer are plotted in Figs.
\ref{fig:EXIT2_DFE_LE_10dB} and \ref{fig:EXIT3_BiDFE_13dB}. They
describe the EXIT charts corresponding to $\mathbf{h_2}$ at a 10 dB
SNR when QTI-DFE is used for the main equalizer and the EXIT chart
on $\mathbf{h_3}$ at a 13 dB SNR when TI-BiDFE is used for the
main equalizer, respectively.  As the figures indicate, both SISE
algorithms appear to widen the EXIT chart tunnels with the aid of the branch equalizers while the trajectory of the single main equalizer itself tends to get stuck in the bottleneck regions. 
Accordingly, both SISE algorithms reach the maximum MI value at a considerably smaller
number of steps than the single equalizer scheme, indicating a
faster convergence for the SISE schemes.

\begin{figure}[!t]
\centering
\includegraphics[width=14.0cm]{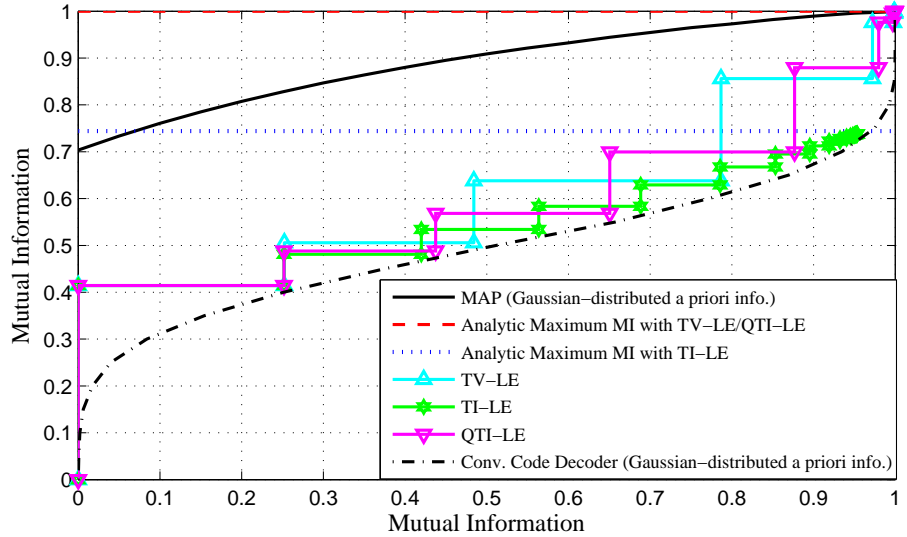}
\caption{EXIT Chart of LE with different filter types on Channel $\mathbf{h_2}$ at SNR=12 dB.}\label{fig:EXIT2_LE_Filter_12dB}
\end{figure}

The EXIT chart of the LE with various filter types on
$\mathbf{h_2}$ at a 12 dB SNR is also shown in Fig.
\ref{fig:EXIT2_LE_Filter_12dB}. As the figure
shows, the trajectory of QTI-LE shows a clear path from 0
bits of MI to 1 bit of MI, in the same way TV-LE does while
TI-LE fails to approach 1 bit of MI even though it keeps moving
up towards its own maximum MI limit as the number of iterations
increases. A word of caution in interpreting Fig.
\ref{fig:EXIT2_LE_Filter_12dB} is in order. The BER performance is not always consistent with the EXIT chart performance;
the MI in the EXIT chart depends on the overall quality
of the extrinsic LLRs and a few erroneously generated extrinsic
LLRs have little effect on the MI, while this is not the case for BER
performance.

\begin{figure}[!t]
\centering
\includegraphics[width=14.0cm]{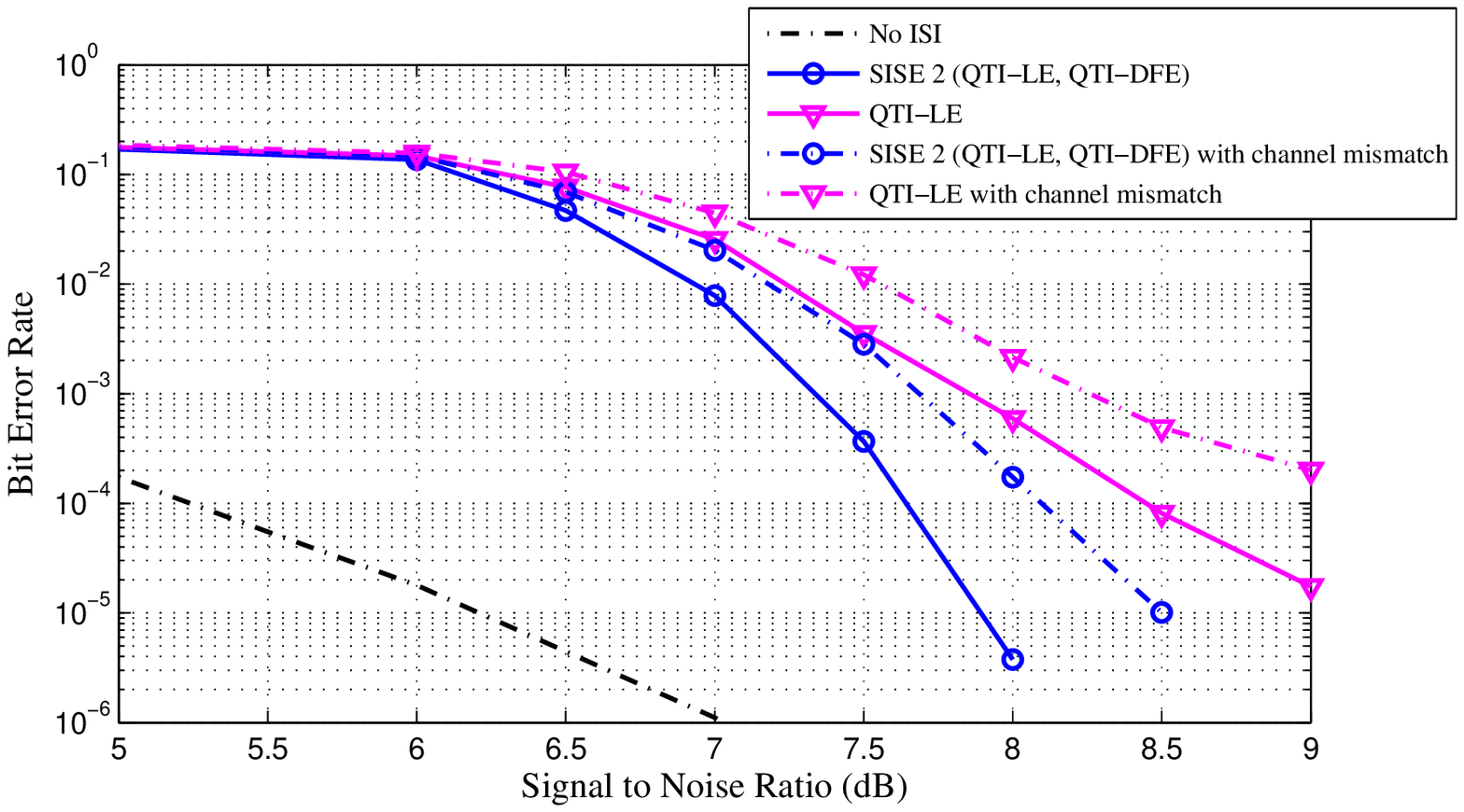}
\caption{LE-based BER curves on Channel $\mathbf{h_4}$ after 20 outer iterations: with and without channel mismatch.}\label{fig:BER4_LE}
\end{figure}

Finally, the performance of 
SISE 2 with a QTI-LE main equalizer and a QTI-DFE branch equalizer is compared with that of 
the stand-alone QTI-LE for the $\mathbf{h_4}$ channel. The results are summarized in Fig. \ref{fig:BER4_LE}. The no-ISI reference curve is also shown.
Also included in the plot are the error rate curves of the same equalizers in the presence of channel mismatch, which show
each scheme's sensitivity to potential channel estimation errors in practice.
Increasing the filter length for QTI-LE did not provide any performance again, so the QTI-LE curves shown in the figure represents the maximum performance potential of the QTI-LE.
The performance gain of the SISE is clearly observed both with and without channel estimation errors. 
In plotting the channel-mismatched curves of Fig. \ref{fig:BER4_LE}, the equalizers operate 
with an erroneous assumption that the channel response is given by  
$\mathbf{\hat{h}_4} = [\hat{h}_0 \: \hat{h}_1 \: \cdots \: \hat{h}_{10} ]$ with $\hat{h}_i = \left\{ 1 + (0.1/\sqrt{SNR})\varepsilon_i \right\} h_i $ where $\varepsilon_i$ is a zero-mean unit-variance Gaussian random variable. In this modeling, the channel estimation error is allowed to drop with increasing SNR. The channel estimation error changes from one channel tap to next, as well as
from one packet transmission to next. The channel errors, however, remain fixed during the turbo/self-equalization process. 

Before ending this section, we remark that
while the work presented in this paper focuses on fixed deterministic channels,
the approach can be applied to a random channel where the channel characteristics 
change at each transmission. Based on the results presented in our paper, it obviously follows that in a random channel, our scheme will have performance advantage if the particular channel realization happens to possess deep nulls and valleys in the signal band whereas the proposed scheme would not show clear performance advantage over stand-alone equalizers whenever the channel realization gives rise to only mild ISI. This was confirmed via our fading channel simulation, although the results are not presented.

\section{Conclusions}\label{sec:Conclusion}
In this paper, we proposed self-iterating soft equalizers which
can be employed in turbo equalization systems to improve
performance in very severe ISI channels. The proposed algorithms
are designed to utilize the extrinsic information of other
serially concatenated suboptimal equalizers by reducing
correlation on the information generated by other equalizers. The
proposed algorithms show robust performance, even when the
constituent suboptimal equalizers are individually weak. The
proposed SISE schemes also provide good BER performance in uncoded
systems.

\newpage

\begin{thebibliography}{1}
\bibitem {Turbo} {C.\ Douillard, M.\ Jezequel, C.\ Berrou, A.\ Picart, P.\ Didier, and A.\ Glavieux,} ``Iterative correction of intersymbol interference: turbo equalization," \textit{European Trans. Telecommunications}, vol. 6, no. 5, pp. 507-511, Sep.-Oct. 1995.

\bibitem {BCJR74} {L.\ Bahl, J.\ Cocke, F.\ Jelinek, and J.\ Raviv,} ``Optimal decoding of linear codes for minimizing symbol error rate," \textit{IEEE Trans. Information Theory}, vol. IT-20, pp. 284-287, Mar. 1974.


\bibitem {SOVA89} {J.\ Hagenauer and P.\ Hoeher,} ``A Viterbi algorithm with soft-decision outputs and its applications," \textit{in Proc. IEEE Global Telecommunications Conference (GLOBECOM)}, Dallas, TX, Nov. 1989, pp. 1680-1686.


\bibitem {Wang99} {X.\ Wang and H.\ V.\ Poor,} ``Iterative (turbo) soft interference cancellation
and decoding for coded CDMA,"  \textit{IEEE Trans. Communications}, vol. 47, no. 7, pp. 1046-1061 , Jul. 1999.

\bibitem {Chan01} {A.\ Chan and G.\ Wornell,} ``A class of block-iterative equalizers for intersymbol
interference channels: fixed channel results,"  \textit{IEEE Trans. Communications}, vol. 49, no. 11, pp. 1966-1976 , Nov. 2001.

\bibitem {Wu01} {Z.\ Wu and J.\ Cioffi,} ``Low complexity iterative decoding with decision-aided equalization for magnetic recording channels,"  \textit{IEEE J. Selected Areas in Communications}, vol. 19, no. 4, pp. 699-708, Apr. 2001.

\bibitem {TSK02} {M.\ T\"uchler, A.\ Singer, and R.\ K\"otter,} ``Minimum mean squared error equalization using \textit{a priori} information,"  \textit{IEEE Trans. Signal Processing}, vol. 50, no. 3, pp. 673-683, Mar. 2002.


\bibitem {TE02} {M.\ T\"uchler, R.\ K\"otter, and A.\ Singer,} ``Turbo equalization: principles and new results,"  \textit{IEEE Trans. Communications}, vol. 50, no. 5, pp. 754-767, May 2002.

\bibitem {Laot01} {C.\ Laot, A.\ Glavieux, and J.\ Labat,} ``Turbo equalization: adaptive equalization and
channel decoding jointly optimized," \textit{IEEE J. Selected Areas in Communications}, vol. 19, no. 9, pp. 1744-1752, Sep. 2001.


\bibitem {Honig04} {M.\ Honig, G.\ Woodward, and Y.\ Sun,} ``Adaptive iterative multiuser decision
feedback detection," \textit{IEEE Trans. Wireless Communications}, vol. 3, no. 2, pp. 477-485, Mar. 2004.


\bibitem {Laot05} {C.\ Laot, R.\ Le Bidan, and D.\ Leroux,} ``Low-complexity MMSE turbo equalization:
a possible solution for EDGE,"  \textit{IEEE Trans. Wireless Communications}, vol. 4, no. 3, pp. 965-974, May 2005.


\bibitem {Sun05} {Y.\ Sun, V.\ Tripathi, and M.\ Honig,} ``Adaptive turbo reduced-rank equalization for MIMO channels," \textit{IEEE Trans. Wireless Communications}, vol. 4, no. 6, pp. 2789-2800, Nov. 2005.



\bibitem {Moon05} {J.\ Moon and F. R.\ Rad,} ``Turbo equalization via constrained-delay APP estimation with decision feedback," \textit{IEEE Trans. Communications}, vol. 53, no. 12, pp. 2102-2113, Dec. 2005.

\bibitem {Rad05} {F. R.\ Rad and J.\ Moon,} ``Turbo equalization utilizing soft decision feedback," \textit{IEEE Trans. Magnetics}, vol. 41, no. 10, pp. 2998-3000, Oct. 2005.



\bibitem {SFE06} {R.\ Lopes and J.\ Barry,} ``The soft-feedback equalizer for turbo equalization of highly dispersive channels," \textit{IEEE Trans. Communications}, vol. 54, no. 5, pp. 783-788, May 2006.

\bibitem {JeongICC10} {S.\ Jeong and J.\ Moon,} ``Turbo equalization based on bi-directional DFE," \textit{in Proc. IEEE International Conference on Communications (ICC)}, Cape Town, South Africa, May 2010.

\bibitem {Jeong10} {S.\ Jeong and J.\ Moon,} ``Soft-in soft-out DFE and bi-directional DFE," \textit{IEEE Trans. Communications}, vol. 59, no. 10, pp. 2729-2741, Oct. 2011.


\bibitem {iterBAD03} {P.\ Supnithi, R.\ Lopes, and S.\ McLaughlin,} ``Reduced-complexity turbo equalization for high-density magnetic recording systems," \textit{IEEE Trans. Magnetics}, vol. 39, no. 5, pp. 2585-2587, Sep. 2003.

\bibitem {BiSFE06} {J.\ Jiang, C.\ He, E.\ Kurtas, and K.\ Narayanan,} ``Performance of soft feedback equalization over magnetic recording channels," \textit{in Proc. Intermag}, San Diego, CA, May 2006, pp. 795.


\bibitem {BiDFE00} {J.\ Balakrishnan and C.\ Johnson, Jr.,} ``Bidirectional decision feedback equalizer: infinite length results," \textit{in Proc. Asilomar Conf. on Signals, Systems, and Computers}, Pacific Grove, CA, Nov. 2001, pp. 1450-1454.

\bibitem {BAD05} {J.\ Nelson, A.\ Singer, U.\ Madhow, and C.\ McGahey,} ``BAD: bidirectional arbitrated decision-feedback equalization," \textit{IEEE Trans. Communications}, vol. 53, no. 2, pp. 214-218, Feb. 2005.


\bibitem {Papke96} {L.\ Papke, P.\ Robertson, and E.\ Villebrun,} ``Improved decoding with the SOVA in a parallel concatenated (turbo-code) scheme," \textit{in Proc. IEEE International Conference on Communications (ICC)}, Dallas, TX, June 1996, pp. 102-106.

\bibitem {Huang06} {C.\ Huang and A.\ Ghrayeb,} ``A simple remedy for the exaggerated extrinsic information produced by the SOVA algorithm," \textit{IEEE Trans. Wireless Communications}, vol. 5, no. 5, pp. 996-1002, May 2006.

\bibitem {JeongGC11} {S.\ Jeong and J.\ Moon,} ``Self-iterating soft equalizer," \textit{in Proc. IEEE Global Telecommunications Conference (GLOBECOM)}, Houston, TX, Dec. 2011.



\bibitem {RDFE92} {S.\ Ariyavisitakul,} ``A decision feedback equalizer with time-reversal structure," \textit{IEEE J. Selected Areas in Communications}, vol. 10, no. 3, pp. 599-613, Apr. 1992.


\bibitem {Brink01} {S.\ ten Brink,} ``Convergence behavior of iteratively decoded parallel concatenated codes," \textit{IEEE Trans. Communications}, vol. 49, no. 10, pp. 1727-1737, Oct. 2001.









\bibitem {BDC} {J.\ Proakis,} \textit{Digital Communications}, 4th edition, New York: McGraw-Hill Higher Education, 2000, pp. 631.









\end{thebibliography}
\end{document}